\documentclass[twocolumn, journal]{IEEEtran}

\IEEEoverridecommandlockouts

\usepackage{float}
\usepackage{stfloats}
\usepackage{mathrsfs}
\usepackage{diagbox}
\usepackage{bm}
\usepackage{color}
\usepackage{graphicx}
\usepackage{amsmath}
\usepackage{amssymb}
\usepackage{algorithm}
\usepackage{algorithmic}
\usepackage{ulem}
\usepackage{lineno}
\usepackage{lettrine}
\usepackage{subfigure}
\usepackage{epstopdf}
\usepackage{stfloats}
\usepackage{color}
\usepackage{graphicx}
\usepackage{amsmath}
\usepackage{amssymb}
\usepackage{algorithm}
\usepackage{algorithmic}
\usepackage{amsmath}
\usepackage{multirow}
\usepackage{booktabs}
\usepackage{array}
\usepackage{amsthm}
\usepackage{stfloats}
\usepackage{caption}
\usepackage{subfigure}
\usepackage{bm}
\usepackage{booktabs}
\usepackage{setspace}
\usepackage{makecell}
\usepackage{cases}
\usepackage{cancel}

\newcommand{\be}{\begin{equation}}
	\newcommand{\ee}{\end{equation}}
\newcommand{\bea}{\begin{eqnarray}}
	\newcommand{\eea}{\end{eqnarray}}
\newcommand{\ba}{\begin{array}}
	\newcommand{\ea}{\end{array}}
\newcommand{\nid}{\noindent}

\allowdisplaybreaks[4]

\title{A Novel Joint Angle-Range-Velocity Estimation Method for MIMO-OFDM ISAC Systems
	\thanks{Z. Xiao and M. Li are with the School of Information and Communication
		Engineering, Dalian University of Technology, Dalian 116024, China (e-mail:
		xiaozichao@mail.dlut.edu.cn; mli@dlut.edu.cn).}
    \thanks{R. Liu and A. L. Swindlehurst are with the Center for Pervasive Communications and Computing, University of California, Irvine, CA 92697, USA (e-mail: rangl2@uci.edu; swindle@uci.edu).}
	\thanks{Q. Liu is with the School of Computer Science and Technology, Dalian University
		of Technology, Dalian 116024, China (e-mail: qianliu@dlut.edu.cn).}}

\author{Zichao Xiao,
	Rang Liu,~\IEEEmembership{Member,~IEEE,}
	Ming Li,~\IEEEmembership{Senior Member,~IEEE,}
	  Qian Liu,~\IEEEmembership{Member,~IEEE,}\\
        and A. Lee Swindlehurst,~\IEEEmembership{Fellow,~IEEE}
}
\begin{document}
\maketitle
\thispagestyle{empty}
\begin{abstract}
Integrated sensing and communication (ISAC) is emerging as a key technique for next-generation wireless systems.
In order to expedite the practical implementation of ISAC in pervasive mobile networks, it is crucial to have widely deployed base stations with radar sensing capabilities. Thus, the utilization of standardized multiple-input multiple-output (MIMO) orthogonal frequency division multiplexing (OFDM) hardware architectures and waveforms is pivotal for realizing seamless integration of effective communication and sensing functionalities. In this paper, we introduce a novel joint angle-range-velocity estimation algorithm for MIMO-OFDM ISAC systems. This approach exclusively depends on the format of conventional MIMO-OFDM waveforms that are widely adopted in wireless communications. Specifically, the angle-range-velocity information of potential targets is jointly extracted by utilizing all the received echo signals within a coherent processing interval (CPI). The proposed joint estimation algorithm can achieve larger signal-to-noise-ratio (SNR) processing gains and higher resolution by fully exploiting the echo signals and jointly estimating the angle-range-velocity information. A theoretical analysis for maximum unambiguous range, resolution, and SNR processing gains is provided to verify the advantages of the proposed joint estimation algorithm. Finally, the results of extensive numerical experiments are presented to demonstrate that the proposed joint estimation approach can achieve significantly lower root-mean-square-error (RMSE) performance for angle/range/velocity estimation for both single- and multi-target scenarios.
\end{abstract}
	
\begin{IEEEkeywords}
Integrated sensing and communication (ISAC), multiple-input multiple-output orthogonal frequency division multiplexing (MIMO-OFDM), parameter estimation.
\end{IEEEkeywords}
	
\section{Introduction}
	
Next-generation wireless systems are expected to develop beyond traditional communication services and further facilitate a series of innovative applications such as intelligent transportation, manufacturing, healthcare, etc.
These emerging applications not only impose higher demands on communication performance but also require more robust sensing capabilities.
In addition, with the exponential growth of wireless devices and communication demands, spectral resources are becoming increasingly scarce.
Radar frequency bands that occupy large portions of the available spectrum are therefore regarded as a promising choice for communication usage.
From a technical point of view, the technology trend of joint wireless communication and radar sensing is highly self-consistent, as they both seek the use of higher frequencies, wider bandwidths, larger antenna arrays, more attention to line-of-sight channels, and distributed dense deployments. Thus, wireless communication and radar sensing exhibit increasing commonality in system design, hardware platforms, signal processing, etc., which provides a strong motivation for integrating 
their functionalities.
Owing to these factors, integrated sensing and communication (ISAC) has emerged, which focuses on the coexistence, cooperation, and co-design of communication and sensing systems. ISAC has been recognized as a key enabling technology for sixth generation (6G) wireless systems \cite{ITU future prospective2} and has aroused extensive research attention from both academia and industry \cite{Le Zhang overview}-\cite{F Liu overview2}.
	
Many approaches for ISAC have been proposed, and these approaches can be generally categorized as based on either traditional radar system design, traditional communication system design, or dual-functional designs that require special customization.
Radar-based ISAC systems focus on embedding communication symbols into existing radar sensing signals, e.g., linear frequency modulated continuous wave (LFMCW) \cite{Chirp based 2} or frequency-hopping (FH) radar \cite{FH radar based 2}.
Communication-based ISAC systems rely on existing communication hardware architectures and waveforms to perform dual-functional tasks \cite{80211 2}-\cite{PMN 3}.
The third category of dual-functional systems are not restricted to current radar/communication infrastructure or waveforms \cite{Unrestricted Architecture 2}, \cite{Unrestricted Architecture 3}.
Given the ubiquitous availability of wireless communication networks, communication-centric designs are the most likely to facilitate the practical development of ISAC, and thus it is critical to investigate the use of communication transceiver architectures and waveforms to empower wireless networks with sensing capabilities.

In existing commercial wireless communication networks, orthogonal frequency division multiplexing (OFDM) has been widely adopted as the dominant waveform type.
OFDM benefits from its ability to overcome frequency selective fading to achieve high spectral efficiency. In addition, OFDM provides satisfactory radar sensing performance by harnessing frequency diversity to enhance target detection \cite{OFDM radar for detection 1} \cite{OFDM radar for detection 2} and parameter estimation \cite{OFDM radar for estimation}. By exploiting the cyclic prefix (CP) signal structure, OFDM can entirely eliminate inter-range-cell interference \cite{IRCI OFDM radar1}, \cite{IRCI OFDM radar3}. Due to the above, OFDM has been recognized as an attractive practical candidate for realizing ISAC.
Since OFDM is communication-oriented, how to realize high-performance radar sensing functionality based on existing OFDM communication systems is a crucial task for facilitating practical ISAC applications.

Many researchers have explored OFDM waveform design and echo signal processing algorithms for ISAC systems, in which the dual-functional OFDM waveform is optimized to simultaneously perform single/multiple-user communications and target detection/estimation/tracking.
In particular, the seminal work in \cite{C Sturm 3} presented a novel algorithm for estimating the range and velocity of potential targets using OFDM waveforms. High-resolution estimation methods have been presented in \cite{Super Resolution method 1}-\cite{Super Resolution method 3}, and a deep-learning algorithm for terahertz systems was developed in \cite{THz OFDM ISAC}.
In addition, subcarrier power allocation has been investigated \cite{SISO OFDM waveform1}-\cite{SISO OFDM waveform3} to achieve a better performance trade-off for OFDM ISAC systems.
{While the studies \cite{C Sturm 3}-\cite{SISO OFDM waveform3} above have verified the potential viability of employing OFDM waveforms for ISAC, their scope was limited to exploring scenarios involving only a single-antenna transmitter.}

{Multi-input multi-output (MIMO) architectures with multiple transmit and receive antennas have been widely employed in both communication and radar sensing systems.}
MIMO architectures provide additional spatial degrees-of-freedom (DoFs) that can be exploited to achieve spatial multiplexing, spatial diversity, and beamforming gain for both communication and radar sensing functions \cite{J Li MIMO radar}, \cite{Emil}. Thus, MIMO is regarded as a key component of future ISAC systems.
However, when OFDM ISAC is implemented with multiple transmit antennas, the dual-functional transmit waveforms include random communication symbols that are subsequently mixed in the spatial domain and influence the reflections from the targets. This greatly complicates the data decoding and radar target parameter estimation, and necessitates the use of advanced echo signal processing algorithms in MIMO-OFDM ISAC systems.

To avoid the mixture of signals emitted from different antennas, the authors in \cite{MIMO OFDM ISAC 1} proposed to allocate different subcarriers to each antenna, which makes the subsequent parameter estimation much easier.
However, since the available frequency resources are not fully exploited, it is obvious that this method will significantly reduce the communication capacity.
Later, \cite{PMN 1} proposed a compressed sensing (CS)-based method for tackling radar parameter estimation using typical MIMO-OFDM communication signals, but this approach is usually computationally prohibitive.
More recently, \cite{MIMO OFDM ISAC 2} and \cite{MIMO OFDM ISAC 3} introduced a novel estimation strategy that involves first estimating the target angle information and then extracting the target range and velocity.
{In particular, the authors of \cite{MIMO OFDM ISAC 2} employed multiple signal classification (MUSIC) for angle estimation and a two-dimensional (2D) discrete Fourier transform (DFT) for range and velocity extraction.
Although this MUSIC-based approach can provide high-resolution angle estimation, it is complex to implement and its perfomrance degrades in scenarios with a large number of targets or low signal-to-noise ratio (SNR).
In \cite{MIMO OFDM ISAC 3}, a suboptimal but efficient approach based on the DFT and cross-correlation is proposed.}
Although this approach is computationally friendly, the estimation of angle and range only exploits the received echoes from one OFDM symbol.
{Compared with traditional radar algorithms that are performed during a coherent processing interval (CPI) covering multiple OFDM symbols, \cite{MIMO OFDM ISAC 3} suffers from low SNR processing gain and has difficulty handling low SNR scenarios.} 
Moreover, the angle-range-velocity of potential targets are sequentially estimated in both \cite{MIMO OFDM ISAC 2} and \cite{MIMO OFDM ISAC 3}, which inevitably leads to error propagation, 
{and \cite{MIMO OFDM ISAC 3} relies on the cross-correlation (ambiguity function) of the transmitted dual-functional signals, which makes it sensitive to the randomness in the communication information.}

\begin{figure*}[!t]
		\centering
		\includegraphics[width=0.9\linewidth]{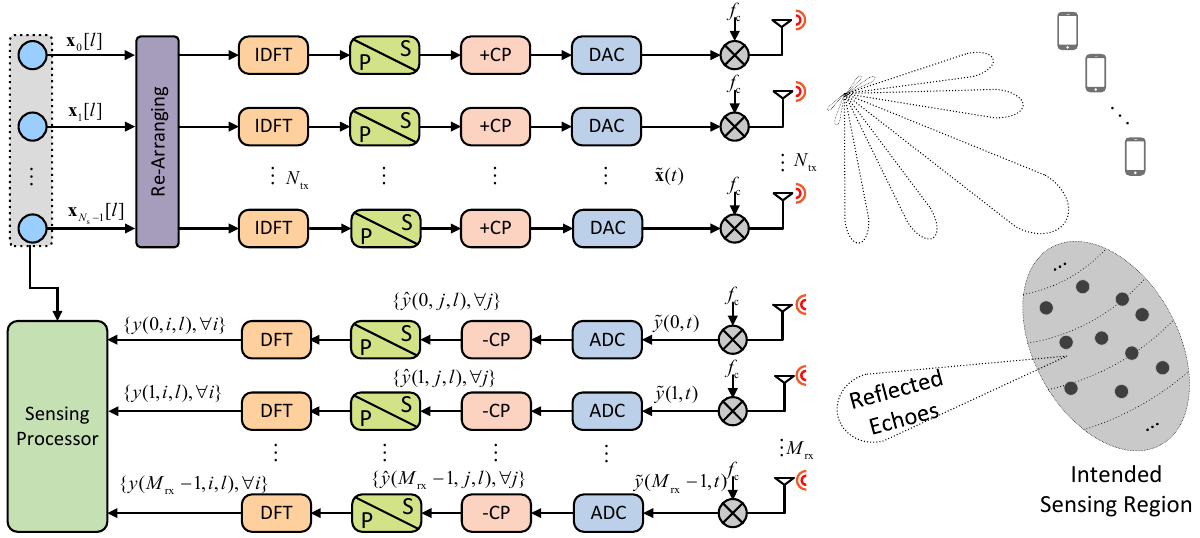}
		\caption{The considered MIMO-OFDM ISAC system.}
		\label{system model fig} 
\end{figure*}

Motivated by the above findings, in this paper we focus on echo signal processing for achieving better parameter estimation performance in MIMO-OFDM ISAC systems.
In particular, we consider a system in which a multi-antenna base station (BS) transmits conventional OFDM waveforms to simultaneously serve multiple communication users and estimate the parameters of multiple point-like targets via processing the received echo signals.
We propose a joint angle-range-velocity estimation approach to fully exploit the available information in the spatial, fast-time (frequency), and slow-time (temporal) domains of the MIMO-OFDM echo signals.
The main contributions are outlined below.
\begin{itemize}
\item First, we propose a novel estimation method for processing the echoes of MIMO-OFDM waveforms to jointly estimate the targets' angle, range, and velocity by fully utilizing all the full three-dimensional (3D) data cube across the spatial, fast-time, and slow-time domains. A DFT-based spectral analysis is first conducted along the spatial dimension to reallocate the signal power according to the angular components. Then, a novel approach is proposed to remove the random communication symbols contained in the cube. Afterwards, spectral analysis along the fast- and slow-time dimensions is performed.
Finally, the angle-range-velocity parameters are jointly estimated via peak finding.
This joint estimation strategy provides substantial SNR processing gains compared with existing approaches that rely on only part of the data cube. In addition, the sensing resolution is significantly improved since the parameters are jointly rather than sequentially estimated.
		
\item Next, we provide a theoretical analysis for the maximum unambiguous range, resolution, and SNR processing gain obtained by the proposed joint estimation approach, from which we gain valuable insights into the achievable sensing performance improvements.
		
\item Finally, simulation results are presented to validate the feasibility and advantages of the proposed joint estimation method.
Compared to existing work, the proposed method provides superior root-mean-squared-error (RMSE) performance and SNR gains for angle-range-velocity estimation.
Moreover, we show that, without sacrificing communication performance, the use of conventional MIMO-OFDM communication waveforms together with the proposed estimation method leads to only a minor sensing performance loss compared to that achieved by standard radar systems employing typical LFMCW waveforms.

\end{itemize}

\textit{Notation}:
Lower-case, boldface lower-case, and upper-case letters indicate scalars, column vectors, and matrices, respectively. The operators
$(\cdot)^T$ and $(\cdot)^H$  denote the transpose and conjugate-transpose operations, respectively, 
$\mathbb{E} \{ \cdot \}$ represents statistical expectation, 
$| a |$ is the magnitude of scalar $a$, 
$\mathbb{C}$ denotes the set of complex numbers, 
and
$\lfloor \cdot \rfloor$ rounds a real number to the nearest integer less than or equal to it.
	
\section{System Model}\label{sec:system model}
	
We consider a mono-static MIMO-OFDM ISAC system as illustrated in Fig. \ref{system model fig}, in which a dual-functional BS equipped with two separate uniform linear arrays (ULAs) of {$N_\mathrm{tx}$} transmit antennas and {$M_\mathrm{rx}$} receive antennas simultaneously performs downlink multi-user communications and radar target parameter estimation.
We assume that the BS operates in full-duplex mode with perfect self-interference (SI) cancellation with the aid of advanced full-duplex techniques \cite{FD 1}-\cite{FD 3}.
Specifically, the BS transmits OFDM waveforms to communicate with $K$ single-antenna users and simultaneously illuminates multiple point-like targets. Meanwhile, the received echo signals are processed to estimate the angle-range-velocity information of potential targets.
	
\subsection{Transmitted Signal Model}
	
In the considered OFDM system, the carrier frequency is $f_\mathrm{c}$, and the wavelength is $\lambda_\mathrm{c}=c/f_\mathrm{c}$, where $c$ denotes the speed of light.
There are $N_\mathrm{s}$-subcarriers with frequency spacing $\Delta f=1/T_\mathrm{d}$, where $T_\mathrm{d}$ is the OFDM symbol duration.
{For the $l$-th OFDM symbol, we denote the dual-functional
baseband signal transmitted on the $i$-th subcarrier as $\mathbf{x}_i[l] \in \mathbb{C}^{N_\mathrm{tx}}$, $i=0, 1, \ldots,N_\mathrm{s}-1$, $l=0, 1,\ldots,L-1$, where $L$ is the frame length of one CPI.}

{As depicted in Fig. \ref{system model fig}, the $N_\mathrm{s}$ $N_\mathrm{tx}$-dimensional frequency-domain baseband signals $\mathbf{x}_i[l]$ collected in different subcarriers are rearranged into $N_\mathrm{tx}$ $N_\mathrm{s}$-dimensional vectors collected from different transmit antennas. 
Then, $N_\mathrm{s}$-point inverse DFT (IDFT) processors are utilized to transform these frequency-domain vectors to the time domain.
These signals are then arranged serially in chronological order, and 
an $N_\mathrm{cp}$-point CP of duration $T_\mathrm{cp}$ is inserted to avoid inter-symbol interference (ISI) for both communication and sensing.}
The CP length should be greater than the length of the channel impulse response to avoid ISI for downlink communications.
In order to eliminate ISI at the sensing receiver, the CP duration should also be larger than the roundtrip delay between the BS and the furthest target.
{After conversion to analog, the baseband signal is expressed as}
\begin{equation}\label{transmitted signal}
{
		\widetilde{\mathbf{x}}(t)\triangleq \sum_{i=0}^{N_\mathrm{s}-1} \sum_{l=0}^{L-1} \mathbf{x}_i[l] e^{\jmath 2\pi i \Delta f t} \mathrm{rect} \left(\frac{t-l T}{T} \right),}
\end{equation}
where $T\triangleq T_\mathrm{d}+T_\mathrm{cp}$ is the total symbol duration and $\mathrm{rect} ( {t}/{T})$ denotes a rectangular pulse of duration $T$.
Finally, the baseband analog signal is up-converted to the radio frequency (RF) domain via $N_\mathrm{tx}$ RF chains with carrier frequency $f_\mathrm{c}$ and then emitted through the antennas.
	
\subsection{Communication Signal Model}
	
After propagating through downlink communication channels, the OFDM signals are received by the single-antenna users and then demodulated into communication symbols.
The communication receiver employs a series of operations including down-conversion, analog-to-digital converting (ADC), CP removal, {serial-to-parallel conversion,} and an $N_\mathrm{s}$-point DFT.
For the $k$-th user, the frequency-domain signal on the $i$-th subcarrier during the $l$-th OFDM symbol is written as
\begin{equation}
		y_{i,k}[l]\triangleq\mathbf{h}_{i,k}^H \mathbf{x}_i[l]+z_{i,k}[l],
\end{equation}
where the vector $\mathbf{h}_{i,k}\in \mathbb{C}^{N_\mathrm{tx}}$ denotes the frequency domain channel between the BS and the $k$-th user, and $z_{i,k} \in \mathcal{CN}(0,\sigma_\mathrm{c}^2)$ denotes additive white Gaussian noise (AWGN).

\subsection{Sensing Signal Model}
	
From the radar sensing perspective, the dual-functional BS attempts to estimate the angle-range-velocity information of multiple point-like targets by processing the received echo signals.
We assume that there are $Q$ targets within the area of interest, and the angle-range-velocity information of the $q$-th target is denoted as $\theta_q$, $R_q$, and $v_q$, respectively, $q\in \mathcal{Q} \triangleq \{1,\ldots,Q\}$.
Note that the angle of arrival (AoA) and the angle of departure (AoD) are both equal to $\theta_q$ in the considered mono-static ISAC system.

The emitted signals will first reach the $Q$ targets and then be reflected back to the receive antennas of the BS.
{During this process, signals will experience the relative propagation delay between the transmit/receive antennas, the round-trip propagation delay between the array reference points, and potentially a Doppler frequency shift.
Thus, the baseband echo signal received by the $m$-th antenna can be expressed as
\be	 \label{simplified received baseband echo signal}
\begin{aligned}
			\widetilde{y}(m,t)\triangleq &
			\sum_{q=1}^Q \beta_q
            \mathbf{a}^H(\theta_q)\widetilde{\mathbf{x}}(t-2R_q/c )\\
			&~~~~~~~e^{{-\jmath2\pi m d_\mathrm{r} \sin\theta_q}/{\lambda_\mathrm{c}}}
			e^{\jmath2\pi f_{\mathrm{D},q}t} +  \widetilde{z}(m,t),
	\end{aligned}
\ee
where $m=0,1, \ldots,M_\mathrm{rx}-1$.
In (\ref{simplified received baseband echo signal}), $\beta_q$ is the attenuation coefficient of the $q$-th target with power $ \mathbb{E}\{|\beta_q|^2\} =\sigma_\beta^2$;
$\mathbf{a}(\theta)\triangleq [e^{{\jmath2\pi 0 d_\mathrm{t} \sin\theta}/{\lambda_\mathrm{c}}},e^{{\jmath2\pi 1 d_\mathrm{t} \sin\theta}/{\lambda_\mathrm{c}}},\ldots,e^{{\jmath2\pi (N_\mathrm{tx}-1) d_\mathrm{t} \sin\theta}/{\lambda_\mathrm{c}}}]^T$
is the transmit steering vector for AoD $\theta$, where $d_\mathrm{t}$ represents the transmit antenna spacing;
$d_\mathrm{r}$ is the receive antenna spacing;
$f_{\mathrm{D},q}={2v_q f_\mathrm{c} }/{c}$ indicates the Doppler frequency;
and $\widetilde{z}(m,t)$ denotes independently and identically distributed (i.i.d.) AWGN.
It is generally assumed that the attenuation coefficient and the angle-range-velocity of the targets are constant during one CPI.
Considering that the signal bandwidth is usually much smaller than the carrier frequency, the phase shifts along the spatial axis and the Doppler phase shift within one OFDM symbol are respectively assumed to be identical on all subcarriers.}
In addition, only first-order reflections from the targets are considered due to high attenuation.

{As shown in Fig. \ref{system model fig}, the baseband analog signals (\ref{simplified received baseband echo signal}) are first processed by ADCs with sampling frequency $F_\mathrm{s} \triangleq N_\mathrm{s}\Delta f$, and the CP is removed from the digital samples. The resulting sampled echo signals for the $l$-th symbol-slot can be written as
\be \small
\widehat{y}(m,j, l)\triangleq \widetilde{y}(m,lT+j/{F_\mathrm{s}}+T_\mathrm{cp}), \label{ISI free part}
\ee
where $j=0,\ldots,N_\mathrm{s}-1$ is the sample index.
After serial-to-parallel conversion and application of an $N_\mathrm{s}$-point DFT along the sample index dimension}, the echo signals are finally be obtained. The corresponding mathematical expression for the echo signal received by the $m$-th receive antenna on the $i$-th subcarrier is acquired by substituting expressions (\ref{transmitted signal}) and (\ref{simplified received baseband echo signal}) into (\ref{ISI free part}) and performing the DFT: 
\begin{subequations} \label{established_echo_model}
{
\begin{align}
			y(m,i,l)
			\triangleq&\sum_{q=1}^Q \beta_q  \mathbf{a}^H\left(\theta_q\right)\mathbf{x}_i[l]e^{-\jmath2\pi  m d_\mathrm{r}\sin\theta_q/\lambda_\mathrm{c} }\\
			&\qquad e^{-\jmath 4\pi i\Delta f {R_q}/{c}}
			e^{\jmath 4\pi lT v_q f_\mathrm{c}/c }
			+ z(m,i,l), \nonumber\\
            =&\sum_{q=1}^Q \beta_q  \mathbf{a}^H\left(\theta_q\right)\mathbf{x}_i[l]e^{\jmath m\omega_\mathrm{a}(\theta_q) }e^{\jmath i\omega_\mathrm{r}(R_q)} \label{established_echo_model b}\\
			&	\qquad e^{\jmath l\omega_\mathrm{v}(v_q) }
			+ z(m,i,l), \nonumber
	\end{align}
}
\end{subequations} \vspace{-2.5ex}

\nid {where $z(m,i,l)\sim \mathcal{CN}(0,\sigma_\mathrm{s}^2)$ denotes the DFT of the AWGN. For conciseness, in (\ref{established_echo_model b}) we respectively define the digital frequencies related to the angle, range, and velocity of the $q$-th target as}
	\begin{subequations} \label{frequency definitions}
{		
\begin{align}
			\omega_\mathrm{a}(\theta_q)&\triangleq -2\pi  d_\mathrm{r}\sin\theta_q/\lambda_\mathrm{c},\label{receive spatial frequency}\\
			\omega_\mathrm{r}(R_q)&\triangleq -4\pi \Delta f {R_q}/{c},\label{distance dependent frequency}\\
			\omega_\mathrm{v}(v_q)&\triangleq 4\pi T v_q f_\mathrm{c}/c.\label{velocity dependent frequency}
		\end{align}
}
	\end{subequations}

 \begin{figure*}[!t]
		\centering
		\includegraphics[width=\linewidth]{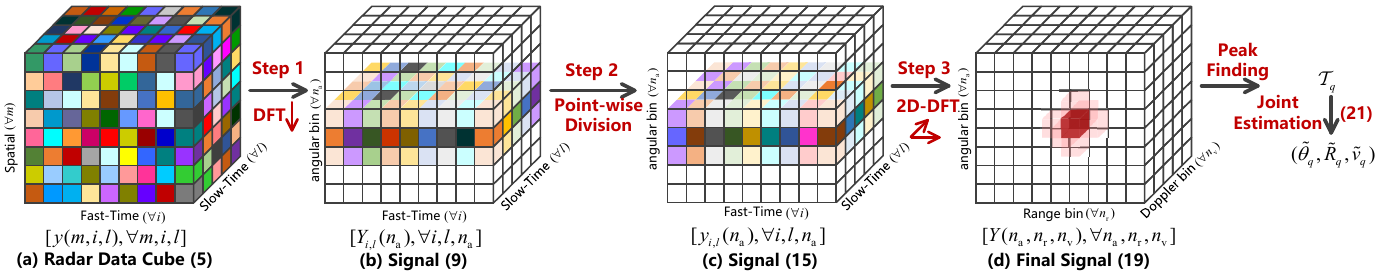}\vspace{-0 ex}
		\caption{{A visualization of the proposed joint estimation process for the single-target scenario.}}
		\label{estimation process} \vspace{-2 ex}
	\end{figure*}

{In this paper, we focus on the radar sensing problem of estimating the angle-range-velocity parameters of the targets based on the received echo signals in (\ref{established_echo_model}) within one CPI.
We see from expression (\ref{established_echo_model b}) that the target angle, range, and velocity are determined by the complex sinusoids of digital frequencies $\omega_\text{r}(\theta_q)$, $\omega_\mathrm{r}(R_q)$, and $\omega_\mathrm{v}(v_q)$, respectively.
However, these sinusoidal functions are multiplied by the signal-dependent coefficient $\mathbf{a}^H(\theta_q)\mathbf{x}_i[l]$, which changes with the transmitted signals and the target AoDs to be estimated.
Thus, the target parameters cannot be directly extracted by performing spectral analysis on (\ref{established_echo_model}).
In order to tackle this difficulty and improve the parameter estimation performance, we propose a novel joint angle-range-velocity estimation method to remove the influence of the signal-dependent coefficients and fully exploit the received echo signals during one CPI.}

\section{Joint Angle-Range-Velocity Estimation} \label{sec:estimation approach}
	
In this section, we focus on extracting the angle-range-velocity information of potential targets from the echo signals (\ref{established_echo_model}).
A novel joint estimation method is proposed to fully exploit the 3D data cube during one CPI and jointly estimate the target parameters.
{Following traditional radar terminology, we refer to the subcarrier dimension, i.e., $\{i=0,1,\ldots,N_\text{s}-1\}$, as \textit{fast-time} and the symbol slot dimension, i.e., $\{l=0,1,\ldots,L-1\}$, as \textit{slow-time}. The dimension corresponding to the receive antennas, i.e., $\{m=0,1,\ldots,M_\mathrm{rx}-1\}$, is referred to as the \textit{spatial dimension}.}
	
\subsection{Echo Signal Observations} \label{sec: analysis}

{Based on the expression for the echo signals in (\ref{established_echo_model}), we have the following observations.
\begin{itemize}
\item Along the spatial dimension for any given $i,l$, the echo signals can be regarded as the summation of $Q$ complex sinusoids with frequencies $\omega_\mathrm{a}(\theta_q)$ and constant amplitudes.
Thus, the frequencies along the spatial dimension are determined by only the target angle.
\item Along the fast-time dimension for any given $m,l$, the echo signals are composed of $Q$ sinusoids with frequencies $\omega_\mathrm{r}(R_q)$, multiplied by the signal-dependent coefficients $\mathbf{a}^H(\theta_q)\mathbf{x}_i[l]$.
Thus, the digital frequencies along the fast-time dimension are determined by the target range and signal-dependent coefficients.
\item Along the slow-time dimension for any given $m,i$, the echo signals also consist of $Q$ sinusoidal functions of frequencies $\omega_\mathrm{v}(v_q)$, also multiplied by the signal-dependent coefficients $\mathbf{a}^H(\theta_q)\mathbf{x}_i[l]$. As a result, the digital frequencies along the slow-time dimension are determined by the target velocity and signal-dependent coefficients.
\item The signal-dependent coefficient $\mathbf{a}^H(\theta_q)\mathbf{x}_i[l]$ is related to the angle $\theta_q$ to be estimated and the transmitted signal $\mathbf{x}_i[l]$.
In the considered ISAC system, the transmitted dual-functional signal $\mathbf{x}_i[l]$ is embedded with random communication symbols, which means that $\mathbf{x}_i[l]$ will change randomly for different indices $i$ and $l$. These signal-dependent coefficient will cause random fluctuations in the fast- and slow-time dimensions, and thus will significantly hinder the analysis needed to obtain the range and velocity information.
\end{itemize}
}
	
{Based on the above observations, it is clear that the angle-range-velocity information is fused across different dimensions of the data cube, and the signal-dependent coefficients $\mathbf{a}^H(\theta_q)\mathbf{x}_i[l]$ are the major obstacle to extracting the target parameters.
In the following, a novel joint estimation method is developed to extract the angle-range-velocity information coupled with the troublesome signal-dependent coefficient.}
	
\subsection{Step 1: Spectral Analysis along the Spatial Dimension} \label{sec: step1}
	
We first analyze the frequency spectrum of the data cube along the spatial dimension to determine the target angles $\theta_q$. In order to facilitate the analysis, the echo signal (\ref{established_echo_model}) is first re-arranged as
	\begin{equation}
		\label{signal model of angle extraction}
		\begin{aligned}
			y(m,i,l) = \sum_{q=1}^Q \mathcal{A}(q,i,l)e^{\jmath m\omega_\mathrm{a}(\theta_q) } + z(m,i,l),
		\end{aligned}
	\end{equation}
{where $\mathcal{A}(q,i,l)$ does not depend on the spatial index $m$ and is defined as}
	\begin{equation}
		\label{expression A angle}
		\begin{aligned}
			\mathcal{A}(q,i,l) \triangleq  \beta_q   \mathbf{a}^H\left(\theta_q\right)\!\mathbf{x}_i[l]
			e^{\jmath i\omega_\mathrm{r}(R_q)}e^{\jmath l\omega_\mathrm{v}(v_q) }.
		\end{aligned}
	\end{equation}
According to (\ref{signal model of angle extraction}), each sequence $\{y(m,i,l),m=0,\ldots,M_\mathrm{rx}-1\},~\forall i,~\forall l$ along the spatial dimension can be viewed as a sum of noise and $Q$ complex sinusoids with angle-dependent frequencies $\omega_\mathrm{a}(\theta_q)$ and amplitude $\mathcal{A}(q,i,l)$.
{While one can easily obtain the angles of potential targets via spectral analysis by using only one such sequence as in \cite{MIMO OFDM ISAC 3}, this ignores information from other available echo signals.
Furthermore, although the estimated angles could be used to remove the signal-dependent coefficients, the angle estimation error will be propagated and even amplified in the estimation of the other parameters.}
Therefore, instead of directly obtaining angle information, the main purpose of spectrum analysis along the spatial dimension is to capture the characteristics of echo signals in different angular bins.

The angular spectral analysis can be conducted by various algorithms, such as the DFT \cite{DFT book}, MUSIC \cite{MUSIC}, ESPRIT \cite{ESPRIT}, compressed sensing based methods \cite{PMN 1}, \cite{PMN 3}, etc.
Each of these has its strengths and limitations. Detailed comparisons can be found in \cite{Braun DFT vs MUSIC}, \cite{Braun dissertation}.
Due to its superior robustness and versatility, we will use the standard DFT approach to develop the proposed joint estimation method that fully utilizes all received echoes in one CPI.
In particular, by applying an $N_\mathrm{a}$-point ($N_\mathrm{a}\geq M_\mathrm{rx}$) normalized DFT on $\{y(m,i,l),m=0,\ldots,M_\mathrm{rx}-1\}$, we can obtain the sequences $\{Y_{i,l}(n_\text{a}),~n_\text{a}=-\lfloor N_\mathrm{a}/2\rfloor,\ldots,\lfloor N_\mathrm{a}/2\rfloor-1\},~\forall i,~\forall l$, given by
    \be\label{DFT sequence of angle}
	Y_{i,l}(n_{\mathrm{a}})
	= \frac{1}{M_\mathrm{rx}}\sum_{m=0}^{M_\mathrm{rx}-1}y(m,i,l) e^{-\jmath m \widetilde{\omega}_\mathrm{a}(n_{\mathrm{a}})},
	\ee
where the $n_\mathrm{a}$-th frequency component ranging from $-\pi$ to $\pi$ is defined as
	\begin{equation}
		\widetilde{\omega}_\mathrm{a}(n_{\mathrm{a}})\triangleq \frac{2\pi n_{\mathrm{a}}}{N_\mathrm{a}}.
	\end{equation}

{Using the spectral analysis in (\ref{DFT sequence of angle}), each echo signal (\ref{signal model of angle extraction}), composed of $Q$ complex sinusoids with frequency $\omega_\mathrm{a}(\theta_q)$ and amplitude $\mathcal{A}(q,i,l)$, is converted into the summation of $N_\mathrm{a}$ complex sinusoids with frequency $\widetilde{\omega}_\mathrm{a}(n_{\mathrm{a}})$ and amplitude $Y_{i,l}(n_\mathrm{a})$ as
	\begin{equation}
		y(m,i,l) = \sum_{n_\mathrm{a}=-\lfloor N_\mathrm{a}/2\rfloor}^{\lfloor N_\mathrm{a}/2\rfloor-1} Y_{i,l}(n_{\mathrm{a}}) e^{\jmath m\widetilde{\omega}_\mathrm{a}(n_{\mathrm{a}})} \label{each decomposition1}.
	\end{equation}
Thus, the echo signals are extracted into different bins corresponding to the angles of potential targets, as shown in Fig. (\ref{estimation process}b).}
Specifically, the power present in the $n_\mathrm{a}$-th angular bin is $|Y_{i,l}(n_{\mathrm{a}})|^2$, which is related to the probability of the existence of targets whose angle-dependent frequencies $\omega_\mathrm{a}(\theta_q)$ are approximately equal to $\widetilde{\omega}_\mathrm{a}(n_{\mathrm{a}})$.
{Next, we will utilize this characteristic to facilitate removal of the signal-dependent coefficient.}

Let $\mathcal{Q}_{n_\mathrm{a}}$ denote the index set of the targets whose angles are within the $n_\text{a}$-th angular bin, and assume the set has cardinality $|\mathcal{Q}_{n_\mathrm{a}}|=Q_{n_\text{a}}$.
It is obvious that $\mathcal{Q}_{n_\mathrm{a}} \subseteq \mathcal{Q}$, $0\leq Q_{n_\text{a}} \leq Q$, $\mathcal{Q}_{n_\mathrm{a}}\cap \mathcal{Q}_{n_\mathrm{a}'}=\emptyset,~\forall {n_\mathrm{a}}\neq {n_\mathrm{a}'}$, and $\sum_{n_\text{a}}\mathcal{Q}_{n_\mathrm{a}}=\mathcal{Q}$.
For the angles within bin $n_\text{a}$, we have the following approximation
	\begin{equation}
		\omega_\mathrm{a}(\theta_q)  \approx \widetilde{\omega}_\mathrm{a}(n_{\mathrm{a}}), ~~ \forall q\in \mathcal{Q}_{n_\mathrm{a}}. \label{angle approx1}
	\end{equation}
{With the definition (\ref{receive spatial frequency}), the angles of the targets in the $n_\text{a}$-th angular bin can be further approximated as}
\be\label{eq:omegat}
{
\theta_q \approx \mathrm{arcsin}\left(- \frac{ n_{\mathrm{a}} \lambda_\text{c}}{ d_\mathrm{r}N_\mathrm{a} } \right) , ~~ \forall q\in \mathcal{Q}_{n_\mathrm{a}},
}
\ee
{and we define $\theta_{n_\mathrm{a}}\triangleq  \mathrm{arcsin}\left(- { n_{\mathrm{a}} \lambda_\text{c}}/(d_\mathrm{r}N_\mathrm{a}) \right)$.}

{Substituting (\ref{signal model of angle extraction}) and (\ref{expression A angle}) into (\ref{DFT sequence of angle}) and using the approximations in (\ref{angle approx1}) and (\ref{eq:omegat}), the amplitude of the $n_\text{a}$-th angular bin can be approximated as}
	\begin{equation} \label{amplitude of the angular bin}
{
    \begin{aligned}
		Y_{i,l}(n_{\mathrm{a}})\approx &\sum_{q \in \mathcal{Q}_{n_\mathrm{a}}} \mathcal{A}(q,i,l), \\
        \approx &\sum_{q\in\mathcal{Q}_{n_\text{a}}}\beta_q
\mathbf{a}^H(\theta_{n_\mathrm{a}})\mathbf{x}_i[l]e^{\jmath i\omega_\mathrm{r}(R_q)}e^{\jmath l\omega_\mathrm{v}(v_q) },
    \end{aligned}	
    }
\end{equation}
where the noise component is ignored to focus on extracting the desired parameters from the target echo signals.
{We observe that $Y_{i,l}(n_{\mathrm{a}})$ consists of $Q_{n_\text{a}}$ components, each of which is the signal-dependent coefficient $\mathbf{a}^H(\theta_{n_\mathrm{a}})\mathbf{x}_i[l]$ multiplied by complex sinusoids whose frequencies are related to the range and velocity of the targets.
Since the signal-dependent coefficient changes along both the fast- and slow-time dimensions, it hinders the estimation of range and velocity using standard spectral analysis algorithms.}
Thus, the next step is to eliminate the impact of this signal-dependent coefficient.

\subsection{Step 2: Signal-Dependent Coefficients Removing}\label{remove MS part}

After spectral analysis along the spatial dimension, the amplitude of the echo signals with different angle-dependent frequencies is extracted.
{Then the signal-dependent term $\mathbf{a}^H(\theta_{n_\mathrm{a}})\mathbf{x}_i[l]$ is removed to eliminate its influence on the spectral analysis along the fast- and slow-time dimensions to extract the estimates of range and velocity. Since the transmitted signals $\mathbf{x}_i[l]$ are known at the dual-functional BS, it would be straightforward to use this informtion together with the estimated target angle bins to remove the signal-dependent term $\mathbf{a}^H(\theta_{n_\mathrm{a}})\mathbf{x}_i[l]$, as is done in \cite{MIMO OFDM ISAC 2} and \cite{MIMO OFDM ISAC 3}. However, only part of the 3D data cube is used in this sequential strategy, and in addition, the angle estimation errors will propagate and degrade the subsequent range and velocity estimation.
Moreover, the method in \cite{MIMO OFDM ISAC 3} is sensitive to the ambiguity function associated with the transmitted signals, and its performance degrades due to the randomness of the communication symbols. 
To avoid these drawbacks, we propose to completely remove the signal-dependent term for each angular bin instead of only focusing on the few estimated target angule bins associated with potential targets.
Thus, our approach performs a joint angle-range-velocity estimation to enhance performance by fully exploiting all echo signals during each CPI.}

From equation (\ref{amplitude of the angular bin}), we observe that the signal-dependent term $\mathbf{a}^H(\theta_{n_\mathrm{a}})\mathbf{x}_i[l]$ could be directly removed by point-wise division.
However, since the magnitude of the signal-dependent term is not identical for different angular bins, directly dividing $Y_{i,l}(n_\text{a})$ by the signal-dependent coefficient will destroy the properties of the echo signals in the spatial dimension. In particular, the spatial characteristics of the echo signals are changed after dividing $Y_{i,l}(n_\text{a})$ by different amplitude values, which will deteriorate the 
angle estimation performance.

To eliminate the impact of the signal-dependent term, we modify the data to maintain the spatial characteristics of the echo signals, and not change the relationship between the powers of different angle bins.
For this purpose, we introduce a scaling factor $\alpha_{n_\text{a}}$ for the $n_\text{a}$-th angular bin and propose to modify $Y_{i,l}(n_\text{a})$ as follows:
\be  \label{signal of step2}
{y_{i,l}(n_{\mathrm{a}}) \triangleq
\begin{cases}
\frac
{Y_{i,l}(n_{\mathrm{a}})}
{ \alpha_{n_{\mathrm{a}}} \mathbf{a}^H(\theta_{n_\mathrm{a}})\mathbf{x}_i[l]}, ~&\mathrm{if} ~
\mathbf{a}^H(\theta_{n_\mathrm{a}})\mathbf{x}_i[l] \neq 0,\\
Y_{i,l}(n_{\mathrm{a}}),
~&\mathrm{if}~ \mathbf{a}^H(\theta_{n_\mathrm{a}})\mathbf{x}_i[l] = 0,
\end{cases}
}
\ee
{where the selection is eliminate dividing by zero.}
Then, based on the above discussions, the scaling factor $\alpha_{n_\text{a}}$ is chosen such that the power of the $n_\text{a}$-th angular bin remains unchanged, i.e.,
\be
\sum_{i,l} |y_{i,l}(n_{\mathrm{a}})|^2
=\sum_{i,l}|Y_{i,l}(n_\mathrm{a})|^2,~\forall n_\text{a},
\ee
with leads to the following equation for the scaling factor:
\be \label{eq:scaling factor}
{\alpha_{n_{\mathrm{a}}}
=
\sqrt{
\frac
{\sum_{i,l,\mathbf{a}^H(\theta_{n_\mathrm{a}})\mathbf{x}_i[l] \neq 0}
\left|\frac{Y_{i,l}(n_\mathrm{a})}
{\mathbf{a}^H(\theta_{n_\mathrm{a}})\mathbf{x}_i[l]
}\right|^2 }
{\sum_{i,l,\mathbf{a}^H(\theta_{n_\mathrm{a}})\mathbf{x}_i[l] \neq 0}|Y_{i,l}(n_\mathrm{a})|^2}
}.
}
\ee

According to the approximation in (\ref{amplitude of the angular bin}), we can further write each sample in (\ref{signal of step2}) as
\begin{align} \label{concrete signal of step2}
	y_{i,l}(n_{\mathrm{a}})
	\approx
	\sum_{q \in \mathcal{Q}_{n_\mathrm{a}}} \beta_q /\alpha_{n_{\mathrm{a}}}
	e^{\jmath i\omega_\mathrm{r}(R_q)}
	e^{\jmath l\omega_\mathrm{v}(v_q) },
\end{align}
{where the special case with $\mathbf{a}^H(\theta_{n_\mathrm{a}})\mathbf{x}_i[l] = 0$ can be ignored without influencing the result.}
Now we clearly see from equation (\ref{concrete signal of step2}) that the resulting sequences obtained along the fast-time dimension $\{y_{i,l}(n_\text{a}),~\forall i\}$ or the slow-time dimension $\{y_{i,l}(n_\text{a}),~\forall l\}$ are composed of $Q_{n_\text{a}}$ sinusoidal functions whose frequencies are determined by the range $R_q$ or the velocity $v_q$.
Thus, it is straightforward to perform spectral analysis along the two dimensions and jointly estimate the desired angle-range-velocity information, as presented in the next subsection.

\subsection{Step 3: Spectral Analysis along the Fast- and Slow-Time Dimensions and Joint Estimation}\label{Distance and Velocity Extraction part}

\begin{table*}[t]
\caption{{Computational Complexity of Estimation Methods for MIMO-OFDM ISAC Systems}}
\centering
\begin{tabular}{|p{2.2cm}| p{11cm}| }
\hline
Proposed  & $\mathcal{O}\left((\log M_\mathrm{rx}+N_\mathrm{tx}+\log N_\mathrm{s}+\log L +g)M_\mathrm{rx}N_\mathrm{s}L\right)$    \\
\hline
Sequential \cite{MIMO OFDM ISAC 2}  & $\mathcal{O}\left(N_\mathrm{s}L+M_\mathrm{rx})M_\mathrm{rx}^2
+(N_\mathrm{tx}+\log N_\mathrm{s}+\log L+g')QLN_\mathrm{s} \right)$ \\
\hline
Sequential \cite{MIMO OFDM ISAC 3} & $\mathcal{O}\big((\log M_\mathrm{rx}+ g'')M_\mathrm{rx}N_\mathrm{s}L + (N_\mathrm{tx}+ \log N_\mathrm{s}+g'' )QN_\mathrm{s}L + (\log L+g'')QL \big)$ \\
\hline
\end{tabular}
\label{tab: complexity analysis}
\end{table*}

In this section, we show how to employ a 2D-DFT operation along the fast- and slow-time dimensions of the signal in (\ref{signal of step2}) to extract the echo signals into different range and Doppler bins. Then, together with the previously obtained angular spectrum, a joint angle-range-velocity estimation from the three dimensions is proposed.

Specifically, a normalized $(N_\mathrm{r},N_\mathrm{v})$-point ($N_\mathrm{r}\geq N_\mathrm{s},~ N_\mathrm{v}\geq L$) 2D-DFT is implemented on the obtained sequences $\{y_{i,l}(n_\text{a}),~\forall i,~\forall l\}$, yielding the sequences $\{Y(n_{\mathrm{a}},n_{\mathrm{r}},n_{\mathrm{v}}),~n_{\mathrm{r}}=- N_\mathrm{r}+1,\ldots,0,~n_\text{v}=-\lfloor N_\mathrm{v}/2\rfloor,\ldots,\lfloor N_\mathrm{v}/2\rfloor-1\},~\forall n_\text{a}$.
The amplitude $Y(n_{\mathrm{a}},n_{\mathrm{r}},n_{\mathrm{v}})$ can be calculated as
\be \label{2D DFT}
Y(n_{\mathrm{a}},n_{\mathrm{r}},n_{\mathrm{v}})
=
\frac{1}{N_{\mathrm{s}}L}\sum_{i=0}^{N_\mathrm{s}-1}\sum_{l=0}^{L-1}y_{i,l}(n_{\mathrm{a}})e^{-\jmath l \widetilde{\omega}_\mathrm{v}(n_\mathrm{v})}e^{-\jmath i \widetilde{\omega}_\mathrm{r}(n_{\mathrm{r}})},
\ee
where the frequency components of the $n_{\mathrm{r}}$-th range bin and the $n_\text{v}$-th Doppler bin are respectively defined as
\begin{subequations}
\begin{align}
\widetilde{\omega}_\mathrm{r}(n_{\mathrm{r}})&\triangleq \frac{2\pi n_{\mathrm{r}} }{N_\mathrm{r}}, \label{second dimension frequency}\\
\widetilde{\omega}_\mathrm{v}(n_\mathrm{v})&\triangleq \frac{2\pi n_\mathrm{v}  }{N_\mathrm{v}}. \label{third dimension frequency}
\end{align}
\end{subequations}

\begin{algorithm}[!t]
		\caption{Proposed Joint Angle-Range-Velocity Estimator}
		\label{alg}
		\begin{algorithmic}[1]\label{algorithm}
			\REQUIRE $y(m,i,l)$, $N_\mathrm{tx}$, $M_\mathrm{rx}$, $N_\mathrm{s}$, $L$, $N_\text{a}$, $N_\mathrm{r}$, $N_\text{v}$, $d_\text{t}$, $d_\text{r}$, $\lambda_\text{c}$, $c$, $f_\text{c}$, $T$, $\Delta f$, $\mathbf{x}_i[l],~\forall m,i,l$.
			\ENSURE $\theta_q,~R_q,~v_q$.
			\STATE { Perform $N_\text{a}$-point DFT on $y(m,i,l)$ along the spatial dimension to obtain $Y_{i,l}(n_\text{a}),~\forall i,l, n_\text{a}$ in (\ref{DFT sequence of angle}). }			
			\STATE { Calculate scaling factor $\alpha_{n_\text{a}}$ in (\ref{eq:scaling factor}). }
			\STATE { Remove signal-dependent coefficients to obtain $y_{i,l}(n_\text{a})$ in (\ref{signal of step2}). }
			\STATE { Perform 2D-DFT along the fast- and slow-time dimensions to obtain $Y(n_\text{a},n_{\mathrm{r}},n_\text{v})$ in (\ref{2D DFT}). }
			\STATE { Find the peaks of $[Y(n_\text{a},n_{\mathrm{r}},n_\text{v}),~\forall n_\text{a}, n_{\mathrm{r}}, n_\text{v}]$ to obtain the index set $\{ \mathcal{T}_q,\forall q \}$. }
			\STATE { Recover the estimated $\widetilde{\theta}_q$, $\widetilde{R}_q$, and $\widetilde{v}_q$ by (\ref{estimated a-d-v}). }
			\STATE {Return $\theta_q=\widetilde{\theta}_q$, $R_q=\widetilde{R}_q$, and $v_q=\widetilde{v}_q$.}			
		\end{algorithmic}
\end{algorithm}

After the steps described above, the echo signals are extracted into the $N_\text{a}\times N_\mathrm{r} \times N_\text{v}$ angular-range-Doppler bins with the amplitudes given in~(\ref{2D DFT}).
Therefore, we can jointly estimate the targets from these 3D bins via peak finding.
{In particular, we can apply cell-averaging constant false alarm rate (CA-CFAR) processing with a 3D cell to obtain several signal clusters with high amplitudes, and then choose the maximum value of each cluster as the peak, whose 3D coordinate we denote by $\mathcal{T}_q$.}
Finally, the estimated angle-range-velocity of the $q$-th target, defined as  $\widetilde{\theta}_q,~\widetilde{R}_q,~\widetilde{v}_q$, can be recovered using the 3D index $\mathcal{T}_q=\{(n_\text{a},n_{\mathrm{r}},n_\text{v})\}$ as
\begin{subequations}\label{estimated a-d-v}
	\begin{align}
		\widetilde{\theta}_q &= \mathrm{arcsin}\left(- \frac{ n_{\mathrm{a}} \lambda_\text{c}}{ d_\mathrm{r}N_\mathrm{a} } \right) , \label{estimated a}\\
		\widetilde{R}_q&=  -\frac{c n_{\mathrm{r}}}{2 N_\mathrm{r} \Delta f}, \label{estimated d}\\
		\widetilde{v}_q&=  \frac{c n_{\mathrm{v}} }{2 N_\mathrm{v} T f_\mathrm{c}}.\label{estimated v}
	\end{align}
\end{subequations}
{We observe that larger values for $N_\text{a}$/$N_\mathrm{r}$/$N_\text{v}$ yield narrower angular/range/Doppler bins and thus better estimation performance. 
However, considering that the computational complexity of the DFT increases with $N_\text{a}$/$N_\mathrm{r}$/$N_\text{v}$, a proper value for the size of the DFT should be selected. 
Our experiments indicate that  
$N_\text{a}=3M_\text{rx}$/$N_\mathrm{r}=3N_\text{s}$/$N_\text{v}=3L$ provides excellent performance with relatively small computational cost.
In addition, interpolation-based approaches can be employed to compensate for the performance loss due to smaller $N_\text{a}$/$N_\mathrm{r}$/$N_\text{v}$, as discussed in Sec. 3.35 of \cite{Braun dissertation}.}

We emphasize that the angle-range-velocity information is estimated {\it jointly} in the 3D space, rather than estimating these quantities sequentially as in previous approaches, and thus the proposed algorithm achieves better estimation performance, especially when the targets are closely spaced along a particular dimension.

\subsection{Summary and Complexity Analysis} \label{estimation part}

Based on the above descriptions, the proposed joint angle-range-velocity estimation approach is summarized in Algorithm \ref{algorithm} and depicted in Fig. \ref{estimation process}.
In summary, spectral analysis along the spatial dimension is first performed to extract the angular components of the echo signals. Then, a scaling factor is introduced to assist in removing the signal-dependent term in each angular bin without destroying the spatial characteristics of the echo signals.
Afterwards, a 2D-DFT is performed for the fast- and slow-time dimensions to extract the range and Doppler components of the signal returns.
Finally, the angle-range-velocity information is jointly estimated by finding peaks among the obtained 3D bins.

{Next, we provide a brief complexity analysis.
In Step 1, the complexity of the DFT along the spatial dimension is of order $\mathcal{O}(N_\mathrm{s}LN_\mathrm{a}\log M_\mathrm{rx})$ via the Fast Fourier Transform (FFT).
In Step 2, the computational complexity is mainly due to the multiplication $\mathbf{a}^H(\theta_{n_\mathrm{a}})\mathbf{x}_i[l]$, $\forall n_\mathrm{a},i,l$, which is of order $\mathcal{O}(N_\mathrm{tx}N_\mathrm{a}N_\mathrm{s}L)$.
In Step 3, the 2D-DFT requires $\mathcal{O}(N_\mathrm{a}LN_\mathrm{r}\log N_\mathrm{s}+N_\mathrm{a}N_\mathrm{r}N_\mathrm{v}\log L)$ operations.
The complexity of the CA-CFAR-based peak finding is of order $\mathcal{O}(gN_\mathrm{a}N_\mathrm{r}N_\mathrm{v})$, where the value of $g$ is positively and linearly related to the size of the search cells.
Therefore, the total computational complexity of the proposed algorithm is of order $\mathcal{O}\left((\log M_\mathrm{rx}+N_\mathrm{tx}+\log N_\mathrm{s}+\log L +g)M_\mathrm{rx}N_\mathrm{s}L\right)$.
For comparison, the complexity of the state-of-the-art approaches in \cite{MIMO OFDM ISAC 2} and \cite{MIMO OFDM ISAC 3}
is also presented in Table \ref{tab: complexity analysis}, where peak finding factors $g$, $g'$, and $g''$ are linearly related to the size of the cells in the three, two, and one-dimensional searches, respectively.
We can observe that the complexity of the proposed method is related to the size of the 3D data cube, i.e., $M_\mathrm{rx}N_\mathrm{s}L$, since we jointly estimate the target parameters by fully exploiting the received echo signals during one CPI. 
Moreover, our proposed algorithm is more efficient than the method of \cite{MIMO OFDM ISAC 2} which requires high-complexity eigenvalue decomposition operations. 
While the algorithm proposed in \cite{MIMO OFDM ISAC 3} has lower computational complexity, it will suffer from inferior performance since it only utilizes part of the echo signal data cube. 
Numerical results in Sec. \ref{sec:simulation results} will further verify the efficiency of the proposed method.}

\section{Performance Analysis}\label{sec:performance analysis}

In this section we conduct a theoretical analysis to evaluate the performance of the proposed joint estimation algorithm.
In particular, we derive the achievable maximum unambiguous range and resolution achieved by the algorithm, which are two key performance indicators generally considered in conventional radar systems.
Furthermore, the SNR processing gain that representing the ability of the algorithm to overcome the noise is also included.

\subsection{Maximum Unambiguous Range}

To avoid ambiguities, the frequencies related to the angle-range-velocity information should satisfy the following relationships:
\begin{subequations}
\begin{align}
-\pi &\leq\omega_\mathrm{a}(\theta)\leq \pi,\\
-2\pi &\leq\omega_\mathrm{r}(R)\leq0,\\
-\pi &\leq\omega_\mathrm{v}(v)\leq \pi.
\end{align}
\end{subequations}
Assuming these conditions are met, then from (\ref{frequency definitions}) the maximum unambiguous range of the angle, range, and velocity can be respectively calculated as
\begin{subequations}
\begin{align}
\theta_\mathrm{max} &\triangleq \arcsin ( {\lambda_\text{c}}/{2 d_\mathrm{r}}),\\
{R}_{\mathrm{max}}&\triangleq  \frac{c}{2 \Delta f }, \\
{v}_{\mathrm{max}}&\triangleq  \frac{c}{4Tf_\mathrm{c}}.
\end{align}
\end{subequations}
The maximum unambiguous range is determined by the wavelength, the antenna spacing, the frequency spacing, the symbol duration, and the carrier frequency, but is independent of size of the transmit antenna array and the transmit power budget.

\subsection{Resolution}\label{resolution analysis part}

The proposed joint estimation algorithm estimates the targets by distinguishing the angular-range-Doppler bins.
Since these bins are obtained using a DFT, the resolution is determined using DFT analysis \cite{DFT book}.
In particular, the DFT frequency resolution for an $N$-point transform is ${2\pi}/{N}$.
Thus, according to the definitions of the frequencies in (\ref{frequency definitions}), the resolution along the spatial dimension $\Delta_\mathrm{a}$, the fast-time dimension $\Delta_\mathrm{r}$, and the slow-time dimension $\Delta_\mathrm{v}$, respectively, satisfy
\begin{subequations}
	\begin{align}
		\omega_\mathrm{a}(\theta+\Delta_\mathrm{a} )-\omega_\mathrm{a}(\theta) &= 2\pi/M_\mathrm{rx}, \\
		\omega_\mathrm{r}(R+\Delta_\mathrm{r} )-\omega_\mathrm{r}(R) &= 2\pi/N_\mathrm{s},\\
		\omega_\mathrm{v}(v+\Delta_\mathrm{v} )-\omega_\mathrm{v}(v) &= 2\pi/L,
	\end{align}
\end{subequations}
which results in
\begin{subequations} \label{resolutions}
	\begin{align}
		\Delta_\mathrm{a} &= \frac{\lambda_\text{c}}{M_\mathrm{rx}d_\mathrm{r}}, \\
		\Delta_\mathrm{r}  &= \frac{c}{2N_\mathrm{s}\Delta f}, \\
		\Delta_\mathrm{v}  &= \frac{c}{2f_\text{c} L T}.\label{eq: velocity resolution}
	\end{align}
\end{subequations}

Since the proposed algorithm jointly uses the DFT along the spatial, fast-time, and slow-time dimensions, the targets can be separated when any one of the resolution requirements is satisfied.
However, for the existing approaches that sequentially estimate the angle, range, and velocity parameters, the resolution criterion for the first estimation step must be satisfied in order to separate two different targets.
Therefore, the proposed joint estimation algorithm can greatly improve the target resolution.

\subsection{SNR Processing Gain} \label{performance analysis part}

{In addition to the typical metrics of maximum unambiguous range and resolution, robustness to noise is also a critical factor.
To evaluate this, the SNR processing gain, which is the ratio of the output SNR to the input SNR, is analyzed in this subsection.}

Based on the received echo signal model in (\ref{established_echo_model}), the input SNR of the $q$-th target during one CPI can be calculated as
\begin{subequations}\label{received SNR}
	\begin{align}
		\mathrm{SNR}_{\text{i},q}
		\triangleq &\mathbb{E}\Big\{ \sum_{m,i,l}
		\big|
		\beta_q  \mathbf{a}^H(\theta_q)\mathbf{x}_i[l]e^{\jmath m\omega_\mathrm{a}(\theta_q) }
		\\
		&e^{\jmath i\omega_\mathrm{r}(R_q)}
		e^{\jmath l\omega_\mathrm{v}(v_q) }
		\big|^2\Big\}/
		{\mathbb{E}\Big\{\sum_{m,i,l}|z(m,i,l)|^2\Big\}} \nonumber \\
		=&
		\frac{ \sigma_\beta^2  \mathbb{E}\big\{ \sum_{m,i,l}
			|
			\mathbf{a}^H(\theta_q)\mathbf{x}_i[l]
			|^2\big\}}
		{M_\mathrm{rx}N_\mathrm{s}L\sigma_\mathrm{s}^2}\\
		=&\frac{\sigma_\beta^2 \sum_{i,l}|
			\mathbf{a}^H(\theta_q)\mathbf{x}_i[l]
			|^2}
		{N_\mathrm{s}L\sigma_\mathrm{s}^2}.
	\end{align}
\end{subequations}
To derive the output SNR of the proposed algorithm, assume
that the $q$-th target is determined to be in the $(n_\text{a},n_{\mathrm{r}},n_\text{v})$-th angle-range-Doppler bin, i.e., $\left(\omega_\mathrm{a}(\theta_q),\omega_\mathrm{r}(R_q),\omega_\mathrm{v}(v_q) \right)\approx \big(\widetilde{\omega}_\mathrm{a}(n_{\mathrm{a}}),\widetilde{\omega}_\mathrm{r}(n_{\mathrm{r}}), \widetilde{\omega}_\mathrm{v}(n_{\mathrm{v}})\big)$. The amplitude $Y(n_{\mathrm{a}},n_{\mathrm{r}},n_{\mathrm{v}})$ of this bin can be calculated by substituting (\ref{concrete signal of step2}) into (\ref{2D DFT}) as
\begin{equation}
	Y(n_{\mathrm{a}},n_{\mathrm{r}},n_{\mathrm{v}})\approx \beta_q /\alpha_{n_{\mathrm{a}}}. \label{a-d-v approx2}
\end{equation}
Meanwhile, using the same procedure as that for processing the target echo signals in (\ref{DFT sequence of angle}), (\ref{signal of step2}) and (\ref{2D DFT}), the amplitude $Z(n_\mathrm{a},n_{\mathrm{r}},n_\mathrm{v})$ corresponding to the received AWGN $z(m,i,l)$ is given by
\be \label{output noise}
Z(n_\mathrm{a},n_{\mathrm{r}},n_\mathrm{v})\triangleq
\!\!\!\sum_{m,i,l}\!\!
\frac{
	z(m,i,l) e^{-\jmath m \widetilde{\omega}_\mathrm{a}(n_{\mathrm{a}})}e^{-\jmath l \widetilde{\omega}_\mathrm{v}(n_\mathrm{v})}e^{-\jmath i \widetilde{\omega}_\mathrm{r}(n_{\mathrm{r}})}
}
{M_\mathrm{rx}N_{\mathrm{s}}L\alpha_{n_{\mathrm{a}}} \mathbf{a}^H(\theta_{n_\mathrm{a}})\mathbf{x}_i[l] },
\ee
which follows
\begin{equation} \label{distribution2}
	{Z}(n_\mathrm{a},n_{\mathrm{r}},n_\mathrm{v})\sim \mathcal{CN}(0, \sigma_{\text{z},n_\mathrm{a}}^2),
\end{equation}
with noise power
\begin{equation}
	\sigma_{\text{z},n_\mathrm{a}}^2= \sum_{i,l}\frac{ \sigma_\mathrm{s}^2}{M_\mathrm{rx}N_\mathrm{s}^2L^2\alpha_{n_{\mathrm{a}}}^2|\mathbf{a}^H(\theta_{n_\mathrm{a}})\mathbf{x}_i[l]|^2}.
\end{equation}
Thus, the output SNR of the $q$-th target can be calculated as
\begin{subequations}
\begin{align}
\mathrm{SNR}_{\text{o},q} &\triangleq
\frac{\mathbb{E}\big\{ |\beta_q /\alpha_{n_{\mathrm{a}}}|^2\big\}}
{\sigma_{\text{z},n_\mathrm{a}}^2}\\
&\approx
\frac{M_\mathrm{rx}N_\mathrm{s}^2L^2 \sigma_\beta^2}{\sum_{i,l}\sigma_\mathrm{s}^2/
|\mathbf{a}^H(\theta_q)\mathbf{x}_i[l]
			|^2},\label{eq:SNRo}
\end{align}
\end{subequations}
where (\ref{eq:SNRo}) is obtained due to the approximation in~(\ref{eq:omegat}).
Then, the SNR processing gain for estimating the parameters of the $q$-th target is given by
\begin{subequations}
\begin{align}
	\frac{\text{SNR}_{\text{o},q}}{\text{SNR}_{\text{i},q}}
&\approx
\frac{M_\mathrm{rx}N_\text{s}^3L^3/ \sum_{i,l}|
			\mathbf{a}^H(\theta_q)\mathbf{x}_i[l]
			|^2}
{\sum_{i,l}1/|
			\mathbf{a}^H(\theta_q)\mathbf{x}_i[l]
			|^2 } \\
&\leq M_\mathrm{rx}N_\text{s}L, \label{eq: SNRo upper}
\end{align}
\end{subequations}
{where the upper-bound (\ref{eq: SNRo upper}) is derived using the Cauchy-Schwarz inequality, and equality holds if and only if the signal-dependent coefficient has constant modulus, i.e.,  $|\mathbf{a}^H(\theta)\mathbf{x}_i[l]| = \text{constant}$, $\forall i,l$.}
{In summary, by fully utilizing the $M_\mathrm{rx}\times N_\text{s} \times L$-dimensional data cube of the received echo signals, the proposed algorithm can provide up to an $M_\mathrm{rx}N_\text{s}L$-fold improvement in SNR performance. 
}

\section{Numerical Results} \label{sec:simulation results}

In this section, we provide numerical experiments to verify the advantages of the proposed joint estimation method in terms of resolution and robustness to noise.
It should be emphasized that the proposed method is dedicated to processing the echoes of MIMO-OFDM waveforms used in general communication systems.
Therefore, the processing at the communication end is the same as that of conventional communication systems and is
ignored in the following simulations.
Unless otherwise specified, we use the settings based on the 3GPP 5G NR high-frequency standard \cite{3GPP 104}, \cite{3GPP 211}, as listed in Table \ref{tab: parameters}.
{To focus on the evaluation of the proposed algorithm and simplify the transmit waveform design, we consider an ISAC scenario where the communication user is also the target to be sensed}\footnote{{The proposed algorithm can be readily applied to more general ISAC scenarios with separate communication users and sensing targets. In order to achieve better performance in such scenarios, a more sophisticated transmit waveform design should be investigated, which is beyond the scope of this paper.}}.

\begin{table}[!t]
\centering
\begin{small}
\caption{System Settings} \label{tab: parameters}

\begin{tabular}{p{4.5cm} p{1.2cm} p{1.5cm} }
\toprule
Parameter  &Symbol & Value\\
\hline
Carrier frequency       & $f_\text{c}$    &$28$GHz  \\
Subcarrier spacing      &$\Delta f$         & $120$kHz   \\
Number of subcarriers   &$N_\mathrm{s}$     &$512$ \\
OFDM symbol duration    & $T_\mathrm{d}$    & $8.33\mu \mathrm{s}$    \\
CP duration             & $T_\mathrm{cp}$  & $0. 59\mu \mathrm{s}$  \\
Total symbol duration   &$T$                & $8.92 \mu \mathrm{s}$ \\
CPI length  &$L$                &$256$\\
Number of transmit antennas  &$N_\mathrm{tx}$                &$16$\\
Number of receive antennas  &$M_\mathrm{rx}$                &$16$\\
Transmit antenna spacing &$d_\mathrm{t}$    &$0.5 c/f_\text{c}$\\
Receive antenna spacing &$d_\mathrm{t}$     &$0.5 c/f_\text{c}$\\
\bottomrule
\end{tabular} 
\end{small}
\end{table}

The targets are assumed to be randomly located at angles between $[-30^{\circ},30^{\circ}]$, ranges between $[40\mathrm{m},80\mathrm{m}]$, and radial velocities between $[-50\mathrm{m/s},50\mathrm{m/s}]$, {where a negative velocity refers to movement away from the receiving platform. 
We assume a zero-forcing (ZF) precoded communication waveform as follows \cite{ZF}:
\begin{equation}
    \label{precoding process}
		\mathbf{x}_i[l] = \mathbf{W}_i \mathbf{s}_i[l], ~\forall i, l,
\end{equation}
where $\mathbf{W}_i \in \mathbb{C}^{ N_\mathrm{tx} \times K}$ denotes the linear ZF beamformer,
$\mathbf{s}_i[l]\in \mathbb{C}^{ K}$ is the communication symbol vector whose elements are independently drawn from a 16-quadrature amplitude modulation (QAM) constellation.}
In addition to the proposed joint estimation algorithm labeled ``Proposed'', we also include three benchmarks for comparisons.
One is the proposed estimation algorithm without scaling when removing the signal-dependent coefficient (denoted as ``W/o scal.'') to show the importance of maintaining the sum-of-sinusoids model in the spatial domain.
{The other two are the state-of-the-art algorithms proposed in \cite{MIMO OFDM ISAC 2} and \cite{MIMO OFDM ISAC 3}.
These sequential estimation schemes are denoted as ``Sequential \cite{MIMO OFDM ISAC 2}'' and ``Sequential \cite{MIMO OFDM ISAC 3}''.}
{In addition, the performance of a radar system employing a standard LFMCW waveform with the 3D-DFT estimation algorithm is also simulated, and is denoted as ``3D-DFT, LFMCW''. 
The LFMCW waveform does not carry random communication symbols, and is known to exhibit excellent sensing performance given the same time-frequency resources \cite{LFMCW}, e.g., the same power budget, bandwidth, and beamforming pattern as the OFDM waveform.}
The LFMCW waveform emitted by $N_\mathrm{tx}$ antennas within one CPI is given as \cite{LFMCW1}
\begin{equation}\label{LFMCW waveform}\small
\mathbf{x}_{\mathrm{LFMCW}}(t)\triangleq \mathbf{w} \sum_{l=0}^{L-1} e^{\jmath 2\pi (f_\mathrm{c}t+ \frac{N_\mathrm{s}\Delta f (t-lT)^2}{T2})} \mathrm{rect}(\frac{t-lT}{T}),
\end{equation}
where $\mathbf{w}\in \mathbb{C}^{N_\mathrm{tx}\times 1} $ is the beamforming vector obtained using the ZF scheme \cite{ZF} in order to guarantee the same beampattern as that of the OFDM waveform for a fair comparison.
At the receiver side, a standard LFMCW radar receiver with $M_\mathrm{rx}$ antennas uses the classic 3D-DFT estimation method \cite{LFMCW1}-\cite{LFMCW3} to estimate the target parameters.

Both single- and multi-target scenarios are considered to illustrate the parameter estimation performance, which
is evaluated in terms of the RMSE of the estimated angle/range/velocity; for example, for the angle parameter we have
\begin{equation}
	\text{RMSE} \triangleq \sqrt{\mathbb{E}\Big\{\frac{1}{Q}\sum_{q=1}^{Q}(\theta_q-\widetilde{\theta}_q)^2\Big\}},
\end{equation}
where $\theta_q$ is the actual AoA of the $q$-target and $\widetilde{\theta}_q$ is the estimate.

\begin{figure*}	\centering

	\subfigure[Angle estimation performance]{
		\begin{minipage}[t]{0.32\linewidth}
			\centering\includegraphics[width=2.4in,trim=20 0 0 0,clip]{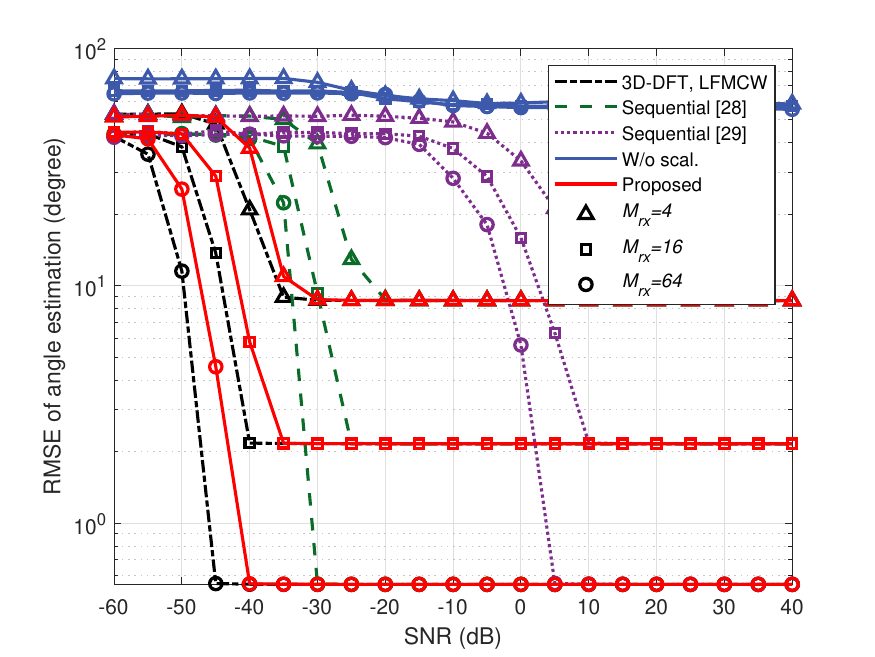}		
    \end{minipage}
	}%
	\subfigure[Range estimation performance]{
		\begin{minipage}[t]{0.32\linewidth}
			\centering
			\includegraphics[width=2.4in,trim=20 0 0 0,clip]{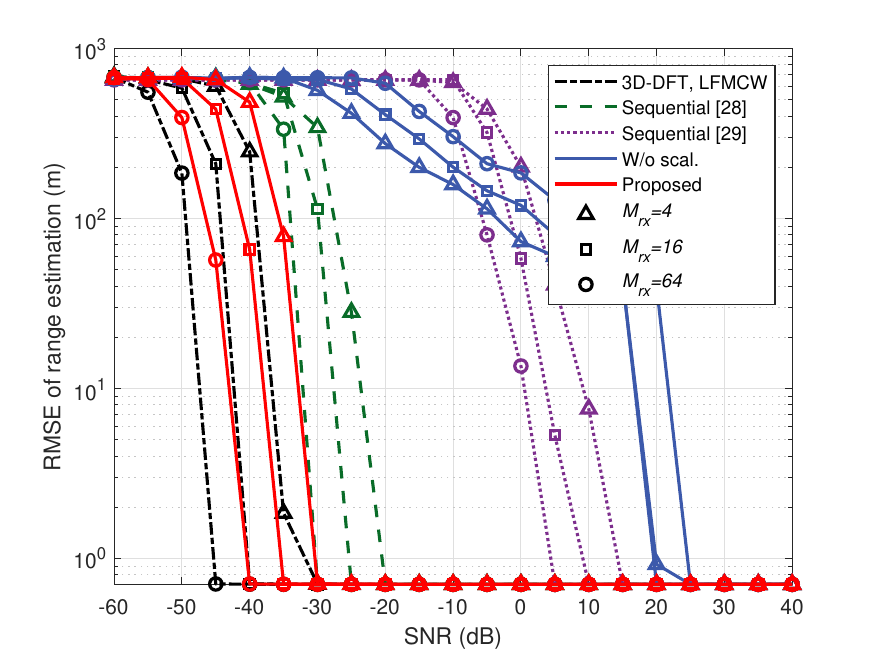}
		\end{minipage}
	}%
	\subfigure[Velocity estimation performance]{
		\begin{minipage}[t]{0.32\linewidth}
			\centering
			\includegraphics[width=2.4in,trim=20 0 0 0,clip]{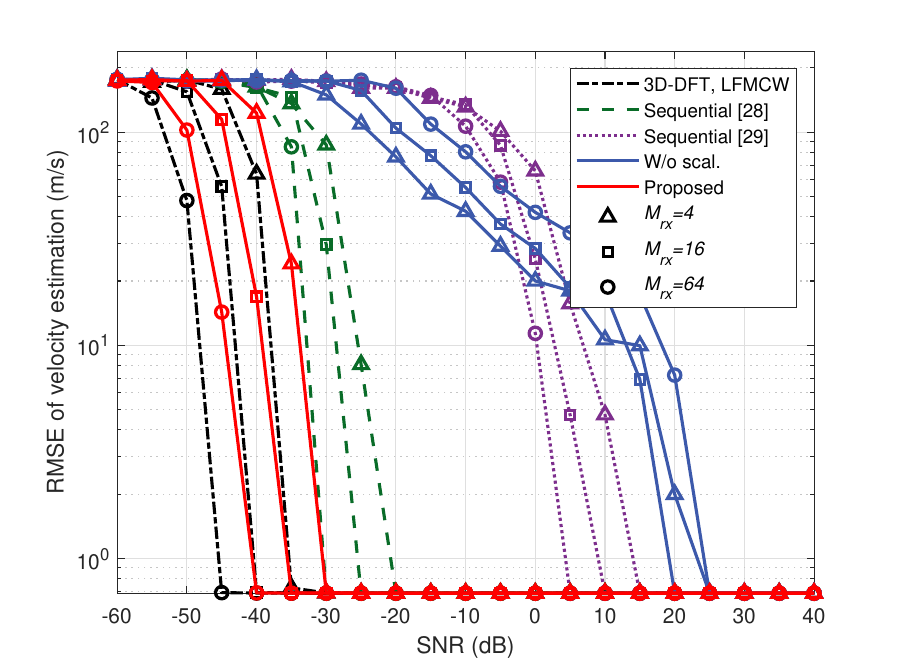}
		\end{minipage}
	}%
	\centering 
	\caption{Estimation performance versus SNR for different numbers of receive antennas, $M_\mathrm{rx}=4,16,64$.}
	\label{fig:RMSE_Nr} 
\end{figure*}
\vspace{-2ex}

\begin{figure*}	\centering
	\subfigure[Angle estimation performance]{
		\begin{minipage}[t]{0.32\linewidth}
			\centering\includegraphics[width=2.4in,trim=20 0 0 0,clip]{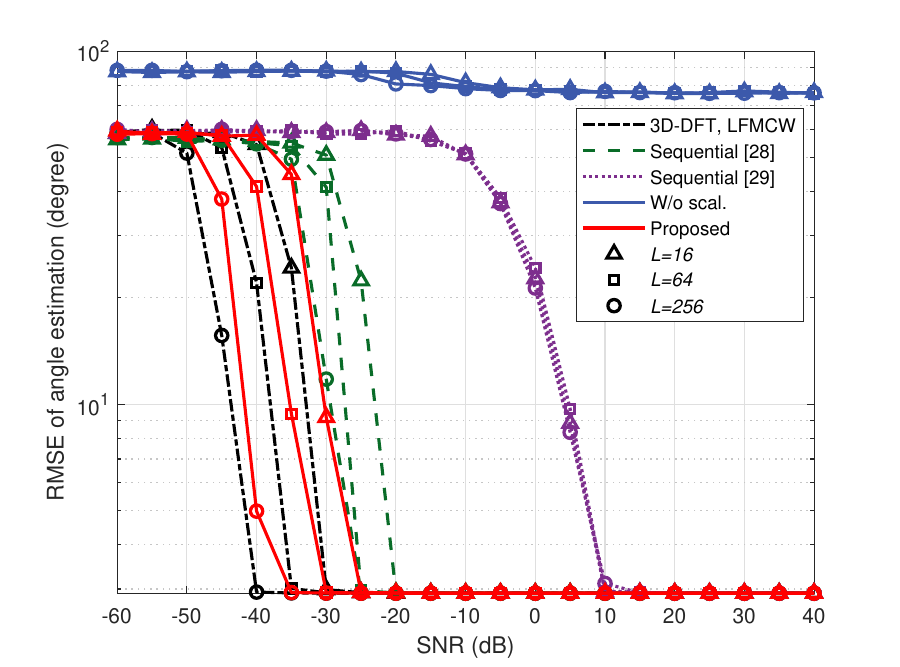}
		\end{minipage}
	}%
	\subfigure[Range estimation performance]{
		\begin{minipage}[t]{0.32\linewidth}
			\centering
			\includegraphics[width=2.4in,trim=20 0 0 0,clip]{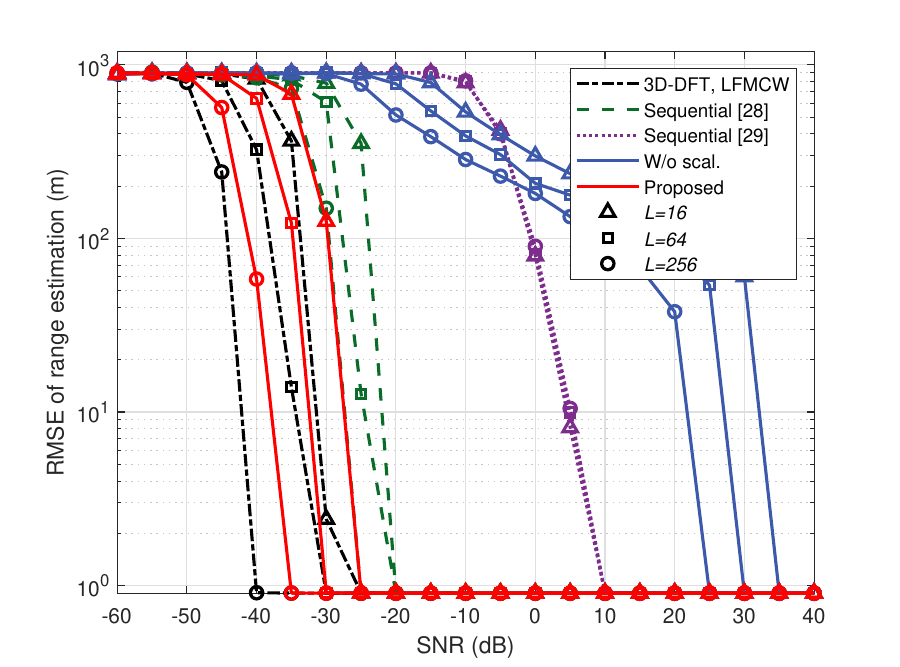}
		\end{minipage}
	}%
	\subfigure[Velocity estimation performance]{
		\begin{minipage}[t]{0.32\linewidth}
			\centering
			\includegraphics[width=2.4in,trim=20 0 0 0,clip]{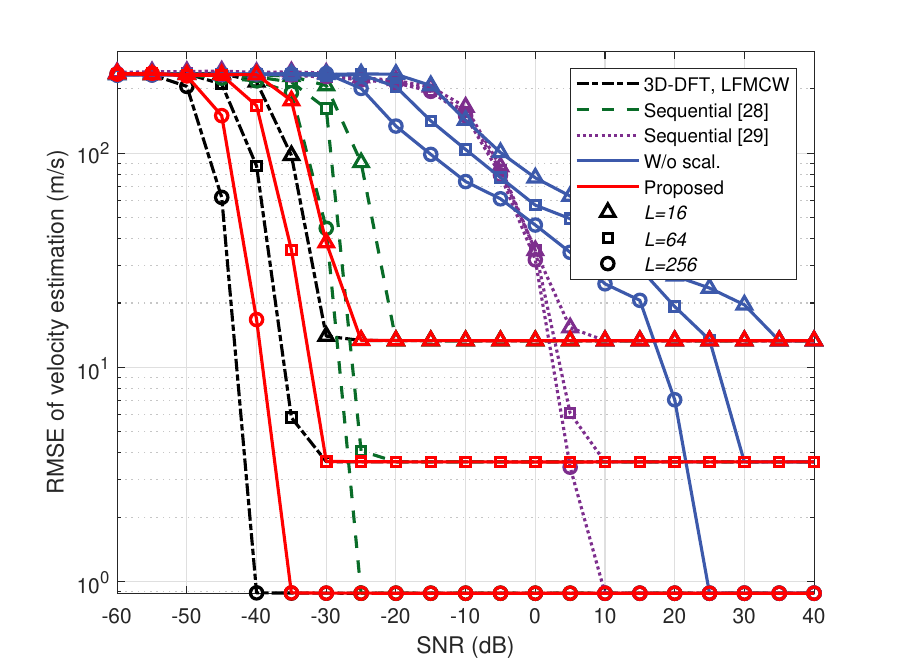}
		\end{minipage}
	}%
	\centering 
	\caption{Estimation performance versus SNR for different CPI lengths, $L=16,64,256$.}
	\label{fig:RMSE_L} 
\end{figure*}

\subsection{Single-Target Scenario}
For the single target case, we set the numbers of angular, range, and velocity bins of all the estimation methods the same, respectively $N_\mathrm{a}=M_\mathrm{rx}$, $N_\mathrm{r}=N_\mathrm{s}$, and $N_\mathrm{v}=L$.
We first show the angle/range/velocity RMSE versus the SNR for different numbers of receive antennas in Fig. \ref{fig:RMSE_Nr}.
Not surprisingly, the estimation error of all schemes decreases as the SNR or the number of receive antennas increases. 
In addition, we observe that the RMSE converges to an error floor at high SNR, since the estimation error is lower bounded by the width of each angular/range/Doppler bin.
The RMSE floor for the angle estimate decreases as the number of receive antennas grows, since a larger number of angular bins leads to a narrower bin width.
Since the spatial characteristics of the echo signals are destroyed when the signal-dependent coefficient is removed, angle estimation for the `W/o scal.' approach is ineffective.
Without our proposed scaling, the range and velocity estimation performance will also be significantly degraded.
Moreover, our proposed joint estimation algorithm significantly outperforms both of the existing algorithms \cite{MIMO OFDM ISAC 2} and \cite{MIMO OFDM ISAC 3}, since we exploit the full 3D data cube. 
Most importantly, the sensing performance obtained using the OFDM communication waveform is very close to that achieved by the LFMCW radar waveform.
This result confirms the advantages of the proposed algorithm and also indicates that the OFDM waveform is an attractive candidate for future ISAC systems.

Next in Fig. \ref{fig:RMSE_L} we show the RMSE performance for different CPI lengths.
As already noted, by fully utilizing the received echo signals, the proposed algorithm achieves the lowest RMSE at each SNR and its performance is very close to the LFMCW radar-only scheme. 
Since only one OFDM symbol is utilized for the angle estimation in \cite{MIMO OFDM ISAC 3}, the estimation performance will be limited for a large $L$.
Specifically, when $L$ increases from $16$ to $256$, the proposed method shows a notable performance improvement of $10\mathrm{dB}$ , while \cite{MIMO OFDM ISAC 2} only improves by about $5\mathrm{dB}$ and \cite{MIMO OFDM ISAC 3} by $0\mathrm{dB}$.

\begin{figure*} \vspace{3ex}
	\centering
	\subfigure[Impact of the number of receive antennas]{
		\begin{minipage}[t]{0.32\linewidth}
			\centering
			\includegraphics[width=2.4in,trim=20 0 0 0,clip]{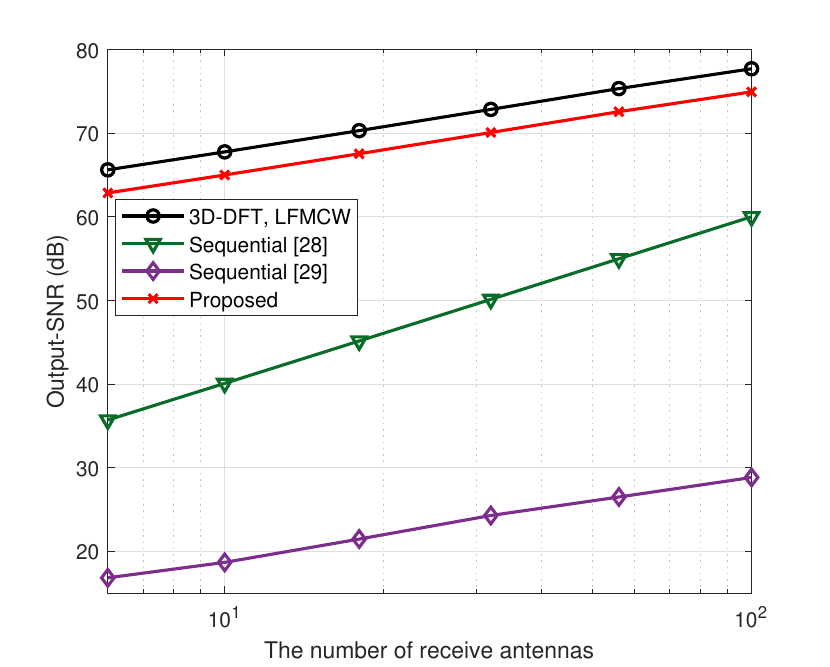}
		\end{minipage}
	}%
	\subfigure[Impact of the number of subcarriers]{
		\begin{minipage}[t]{0.32\linewidth}
			\centering
			\includegraphics[width=2.4in,trim=20 0 0 0,clip]{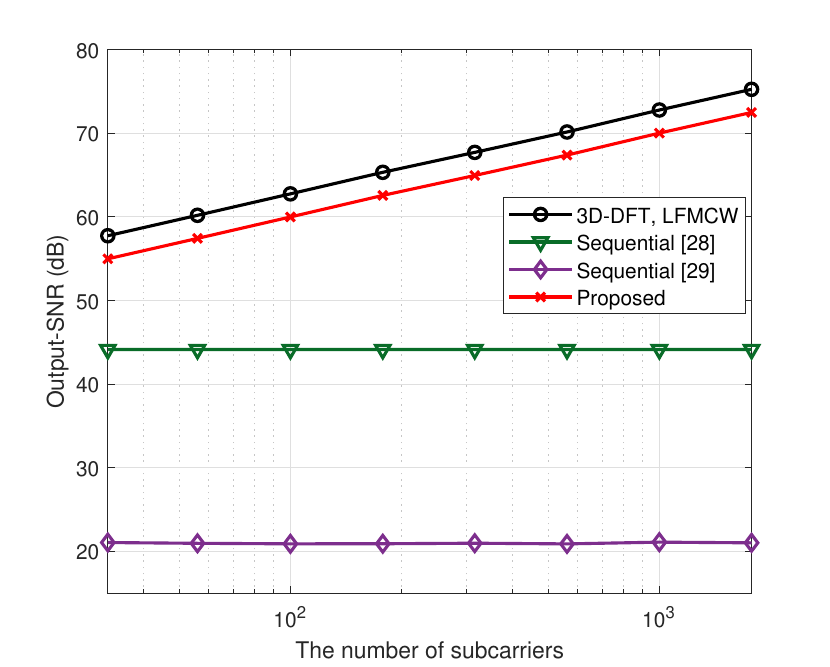}
		\end{minipage}
	}%
	\subfigure[Impact of CPI length]{
		\begin{minipage}[t]{0.32\linewidth}
			\centering
			\includegraphics[width=2.4in,trim=20 0 0 0,clip]{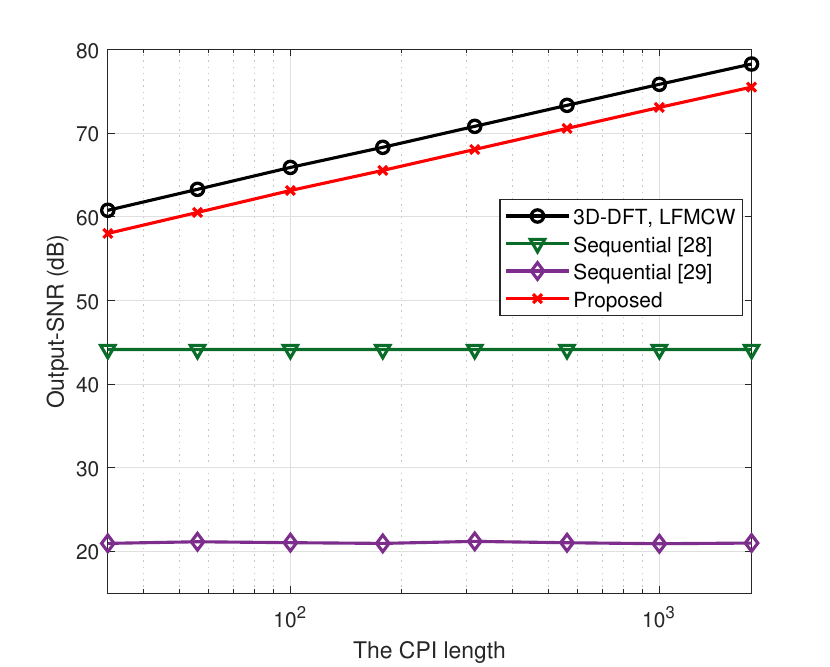}
		\end{minipage}
	}%
	\centering\vspace{1ex}
	\caption{Output-SNR versus the number of receive antennas, subcarriers, and CPI length, respectively. (Input SNR: $10\mathrm{dB}$)} \vspace{0ex}
	\label{fig: SNR}
\end{figure*}

To further quantify the proposed algorithm's robustness to noise, in Fig. \ref{fig: SNR} we plot the output SNR for a $10\mathrm{dB}$ input SNR versus the number of receive antennas, subcarriers, and CPI length, respectively.
Since the algorithm `W/o scal.' cannot provide correct results in this case, results for this approach are not presented in the following simulations.
In addition, given that the performance of the sequential strategy depends mainly on the initial angle estimation results, the output SNR shown for \cite{MIMO OFDM ISAC 2} and \cite{MIMO OFDM ISAC 3} is taken to be the output SNR of the angle estimation only.
We see that the output SNR of the proposed algorithm increases linearly with the number of receive antennas, subcarriers, and the CPI length.
This is consistent with our analytical expression (\ref{eq:SNRo}) and verifies that the proposed joint estimation algorithm can fully exploit the echo signals from all three dimensions.
Comparatively, the output SNR obtained by the sequential estimation algorithms \cite{MIMO OFDM ISAC 2} and \cite{MIMO OFDM ISAC 3} increases only with the number of receive antennas $M_\mathrm{rx}$, since their performance is mainly determined by the first-step angle estimation which is related to $M_\mathrm{rx}$.
We observe that the proposed algorithm has a substantially higher output SNR than its counterparts, about $15\mathrm{dB}$ higher than \cite{MIMO OFDM ISAC 2} and $45\mathrm{dB}$ higher than \cite{MIMO OFDM ISAC 3}, and thus the proposed method is more reliable in handling low SNR situations.
Moreover, we see that the output SNR obtained using the OFDM communication waveform together with the proposed estimation algorithm is comparable to that achieved by the radar-only LFMCW scheme.

\begin{figure}
	\centering
    \subfigure[Impact of the number of transmit antennas]{
		\begin{minipage}[t]{0.47\linewidth}
			\centering
			\includegraphics[width=1.8in,trim=0 0 0 0,clip]{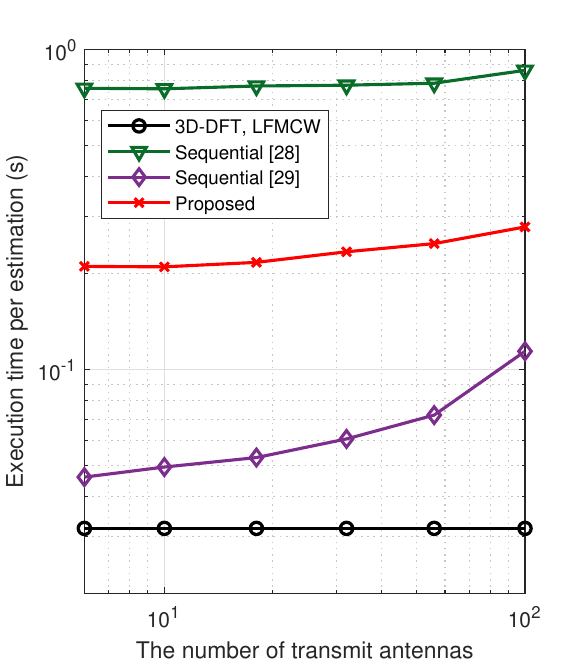}
		\end{minipage}
	}%
	\subfigure[Impact of the number of receive antennas]{
		\begin{minipage}[t]{0.47\linewidth}
			\centering
			\includegraphics[width=1.8in,trim=0 0 0 0,clip]{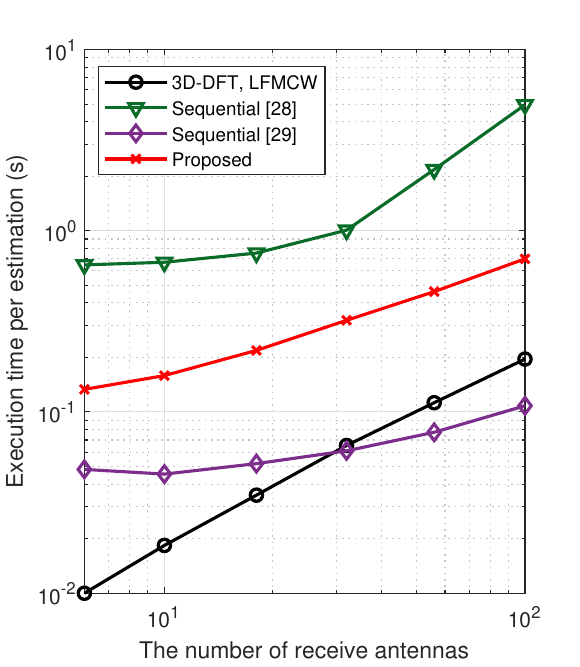}
		\end{minipage}
	} \vspace{0.5 cm}

	\subfigure[Impact of the number of subcarriers]{
		\begin{minipage}[t]{0.47\linewidth}
			\centering
			\includegraphics[width=1.8in,trim=0 0 0 10,clip]{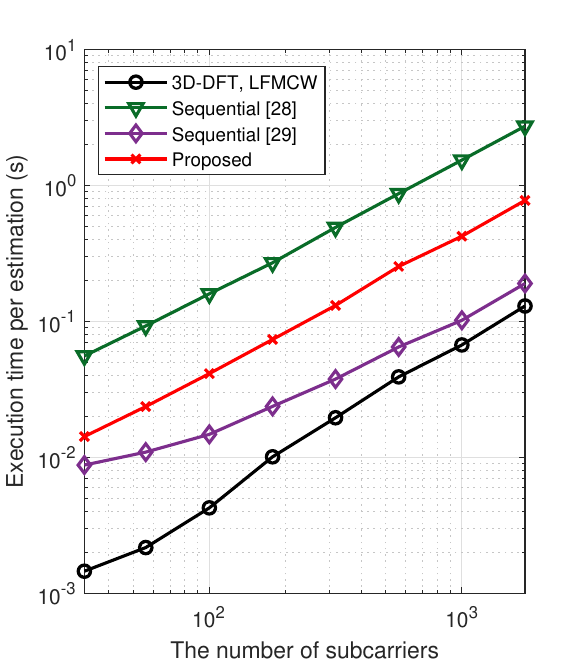}
		\end{minipage}
	}%
	\subfigure[Impact of CPI length]{
		\begin{minipage}[t]{0.47\linewidth}
			\centering
			\includegraphics[width=1.8in,trim=0 0 0 10,clip]{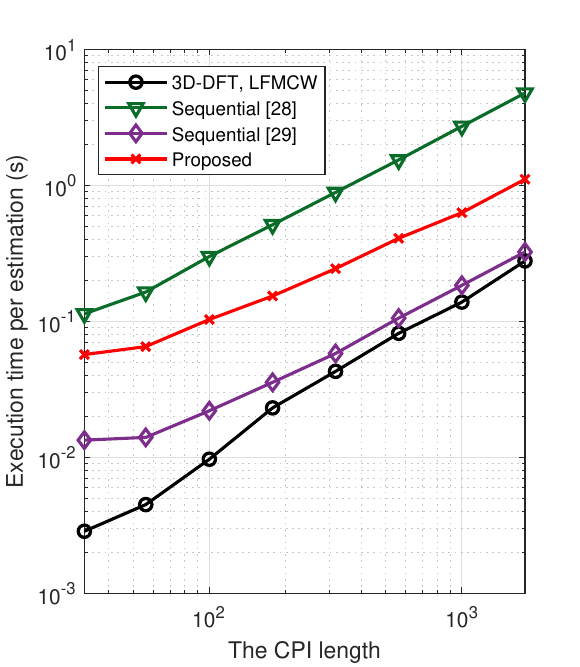}
		\end{minipage}
	}%
	\centering 
	\caption{{Execution time per estimation versus the number of transmit antennas, receive antennas, subcarriers, and CPI length.}}
	\label{fig: time}
\end{figure}

{For practical consideration, in Fig. \ref{fig: time}, we show the execution time per estimation versus the number of transmit antennas, receive antennas, subcarriers, and CPI length. 
We clearly observe that the relationship between the execution times required by different algorithms is consistent with the theoretical analysis in Sec. III-E. 
Specifically, the algorithm in \cite{MIMO OFDM ISAC 3} that only utilizes a portion of the echoes is the most computationally efficient, the algorithm \cite{MIMO OFDM ISAC 2} using eigenvalue decomposition requires the most execution time, and the proposed algorithm lies in between them. 
Although the proposed algorithm is not the most computationally efficient, the required execution time is still acceptable. 
Moreover, the substantial estimation performance improvement and the high achievable communication rate enable the proposed joint estimation algorithm to be a competitive candidate for practical ISAC applications. 
}

\subsection{Multi-Target Scenario} \label{Simulation part: resolution}

{In order to illustrate the estimation performance in terms of resolution, we consider multi-target scenarios in this section.
To ensure satisfactory estimation performance, we set the number of angular bins for the MUSIC-based angle estimation algorithm \cite{MIMO OFDM ISAC 2} as $N_\mathrm{a}=30M_\mathrm{rx}$, and set the number of angular, range, and velocity bins for the other estimation approaches as $N_\mathrm{a}=3M_\mathrm{rx}$, $N_\mathrm{r}=3N_\mathrm{s}$, and $N_\mathrm{v}=3L$. The number of receive antennas is set as $M_\mathrm{rx}=8$.}

\begin{figure*}
	\centering
	\subfigure[Angle estimation performance]{
		\begin{minipage}[t]{0.32\linewidth}
			\centering
			\includegraphics[width=2.4in,trim=10 0 0 0,clip]{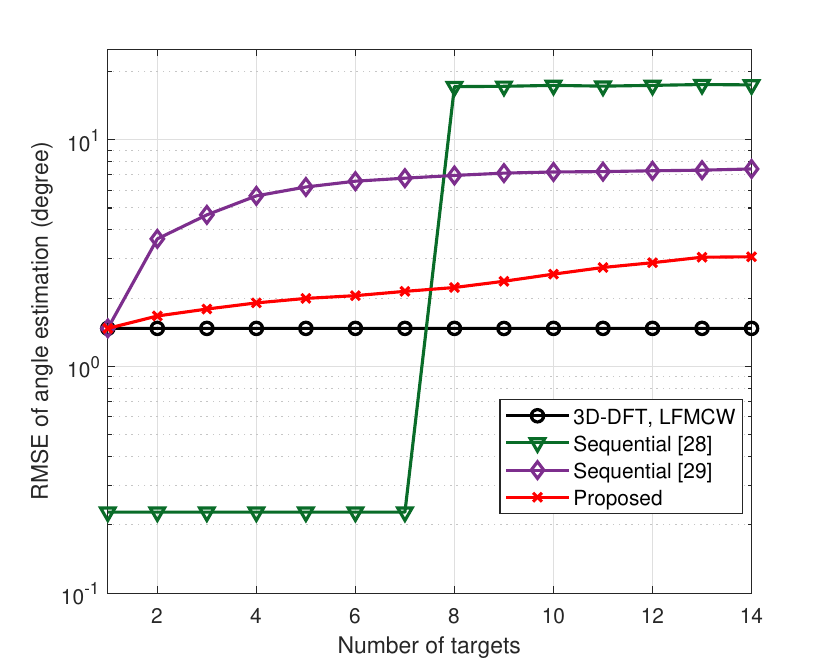}
		\end{minipage}
	}%
	\subfigure[Range estimation performance]{
		\begin{minipage}[t]{0.32\linewidth}
			\centering
			\includegraphics[width=2.4in,trim=10 0 0 0,clip]{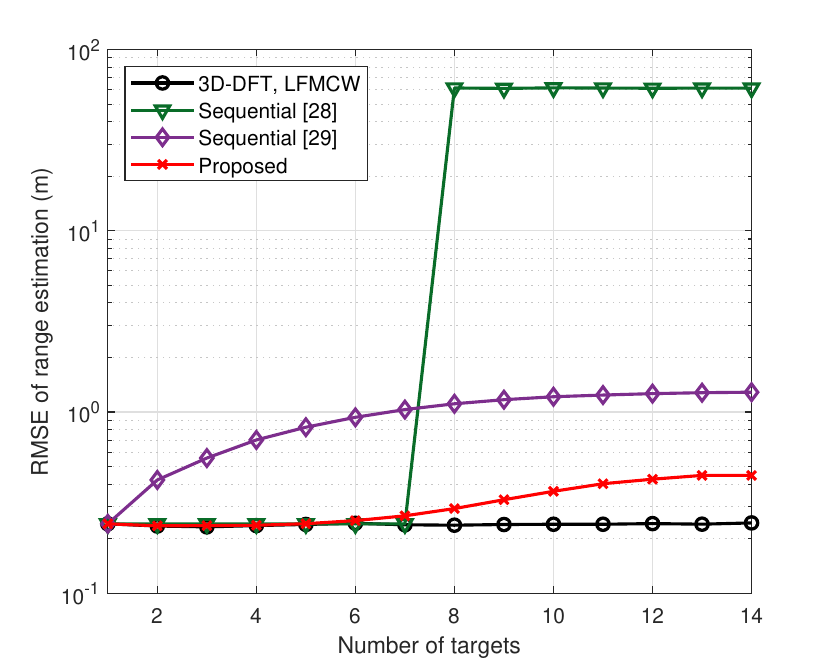}
		\end{minipage}
	}%
	\subfigure[Velocity estimation performance]{
		\begin{minipage}[t]{0.32\linewidth}
			\centering
			\includegraphics[width=2.4in,trim=10 0 0 0,clip]{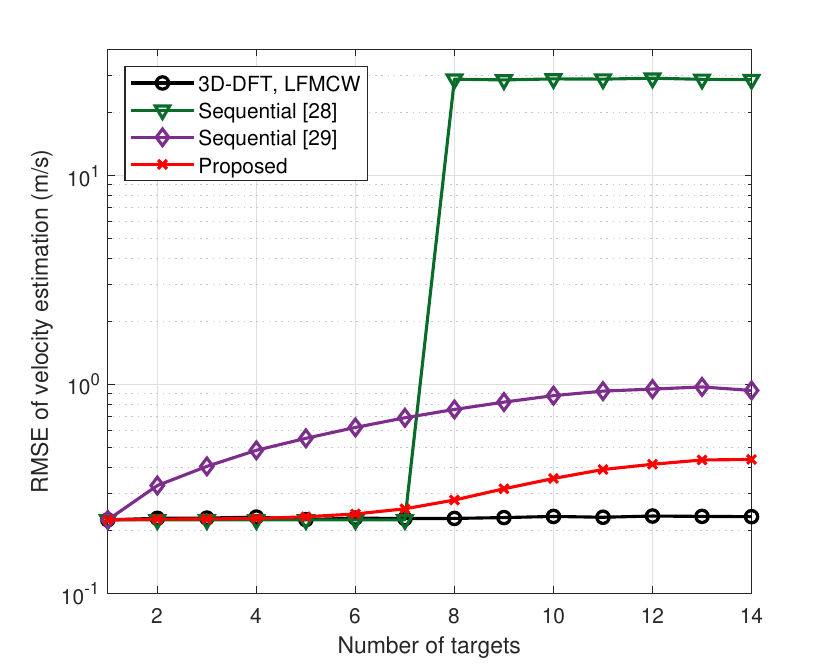}
		\end{minipage}
	}%
	\centering 
	\caption{Estimation performance versus the number of targets, $M_\mathrm{rx}=8$.}
	\label{fig: RMSE_res}
\end{figure*}

{We first present the RMSE versus the number of targets in Fig. \ref{fig: RMSE_res}.
We observe that when the number of targets equals the number of receive antennas, the RMSE of the sequential method \cite{MIMO OFDM ISAC 3} increases quickly, while the proposed method maintains a relatively stable performance thanks to the joint estimation process. 
When the number of targets is greater than the number of receive antennas, the sequential method \cite{MIMO OFDM ISAC 2} no longer provides valid estimates since it is based on the MUSIC method, which requires that the number of angles be less than the number of antennas. 
Although the process of removing the signal-dependent term introduces a certain performance loss, the proposed algorithm using OFDM communication waveforms still achieves performance comparable to the LFMCW radar-only scheme, which provides satisfactory communication and sensing performance for ISAC applications.
}

\begin{figure}[!t]
	\centering
	\includegraphics[width = 3.6 in]{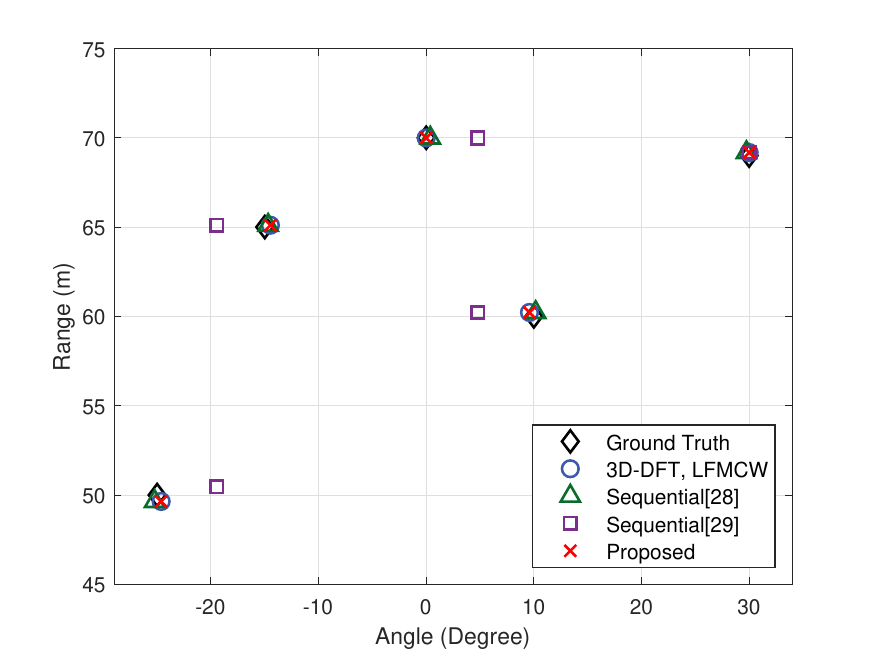}
	\caption{Target estimation results.}
	\label{fig: TF map}
\end{figure}

{In order to obtain better intuition about the performance gap, in Fig. \ref{fig: TF map}
we show the results for a specific case where $Q=5$ point-like targets exist with the angle-range-velocity parameters listed in Table \ref{tab: estimated results}.
We see that the sequential method \cite{MIMO OFDM ISAC 3} exhibits notably worse performance in localizing targets with similar angles because it distinguishes the targets based only on their angle information from the spatial dimension.
Although the method in \cite{MIMO OFDM ISAC 2} also initially observes the data cube from one dimension, it employs the MUSIC method which provides better resolution.
Compared to these two benchmarks, the proposed joint estimation algorithm provides more accurate results close to the LFMCW radar-only scheme using the 3D-DFT. }

\begin{table}[!t]
	\centering
	\begin{small}
		\caption{ Target Parameters}\label{tab: estimated results}
		
		\begin{tabular}{p{1.2cm} p{1.2cm} p{1.2cm}    }
			\hline
			Angle  &Range & Velocity\\
			\hline
			$~~30^{\circ}$   &$69\mathrm{m}$&$-30\mathrm{m/s}$       \\
			$~~~0^{\circ}$&$70\mathrm{m}$&$~~10\mathrm{m/s}$        \\
			$~~~10^{\circ}$&$60\mathrm{m}$&$~~25\mathrm{m/s}$    \\
			$-25^{\circ}$&$50\mathrm{m}$&$~~30\mathrm{m/s}$      \\
			$-15^{\circ}$&$65\mathrm{m}$&$-5\mathrm{m/s}$      \\
			\hline
		\end{tabular}
	\end{small}
\end{table}

\section{Conclusions}\label{sec:conclusions}
In this paper we proposed a novel algorithm for jointly estimating the parameters of multiple targets based on the use of conventional MIMO-OFDM waveforms in an ISAC system.
The algorithm jointly estimates the angle-range-velocity information of potential targets by fully exploiting the received echo signals within a coherent processing interval.
A theoretical analysis for the maximum unambiguous range, resolution, and SNR processing gain was provided to evaluate the performance of the proposed algorithm.
Extensive simulation results verified that the proposed approach can achieve much better parameter estimation performance than the existing algorithms using MIMO-OFDM communication waveforms, as well as performance close to that achieved by a radar-only system using LFMCW waveforms.
{Based on this initial work, we will further investigate algorithm design for ultra-wideband (UWB) MIMO-OFDM waveforms and transmit waveform design for general ISAC scenarios in which the users and targets are in different positions.}

\end{document}